# State-dependent brain responsiveness, from local circuits to the whole brain

**A. Destexhe[1], J Goldman[1], N. Tort-Colet[1], A. Roques[1], J. Fousek[2], S. Petkoski[2], V. Jirsa[2], O. David[2], M. Jedynak[2], C. Capone[3], C. De Luca[3], G. De Bonis[3], P.S. Paolucci[3], E. Mikulan[4], Pigorini[4], M Massimini[4], A. Galluzzi[5], A. Pazienti[5], M. Mattia[5], A. Arena[6], BE Juel[6], E. Hagen[6], J.F. Storm[6], E. Montagni[7], F. Resta[7], F. S. Pavone[7], A. L. Allegra Mascaro[7], A. Dwarakanath[8], TI Panagiotaropoulos[8], J. Senk[9,12], M. Diesmann[9,13,14,15], A. Camassa[10], L. Dalla Porta[10], A. Manasanch[10], M.V. Sanchez-Vives[10,11]**

[1]Paris-Saclay University, Institute of Neuroscience (NeuroPSI), CNRS, Saclay, France; [2]Aix-Marseille University, Marseille, France; [3]INFN, Rome, Italy; [4]UMIL, Milan, Italy; [5]ISS, Rome, Italy; [6]UiO, Oslo, Norway; [7]LENS, Florence, Italy; [8]CEA, NeuroSpin, France; [9]Institute for Advanced Simulation (IAS-6), Jülich Research Centre, Jülich, Germany; [10]Institut d'Investigacions Biomèdiques August Pi i Sunyer (IDIBAPS), Roselló 149-153, 08036 Barcelona, Spain; [11]ICREA, Passeig Lluís Companys 23, 08010 Barcelona, Spain; [12]Sussex AI, School of Engineering and Informatics, University of Sussex, Brighton, United Kingdom; [13]JARA-Institute Brain Structure-Function Relationships (INM-10), Jülich Research Centre, Jülich, Germany; [14]Department of Psychiatry, Psychotherapy and Psychosomatics, School of Medicine, RWTH Aachen University, Aachen, Germany; [15]Department of Physics, Faculty 1, RWTH Aachen University, Aachen, Germany

Present addresses: A. Dwarakanath: Neural Circuits and Cognition, European Institute of Neuroscience, Göttingen 37077, Germany; Cognitive Neuroscience Laboratory, Perception & Plasticity Group, Deutsches Primatenzentrum, Göttingen 37077, Germany. TI Panagiotaropoulos: Department of Psychology, National & Kapodistrian University of Athens, Athens, Greece; Centre for Basic Research, Biomedical Research Foundation of the Academy of Athens (BRFAA), Athens, Greece.

| Figure | Short URL | link |
|---|---|---|
| Figure 2 panel F,G,H | https://t.ly/0kWg2 | https://lab.jsc.ebrains.eu/hub/user-redirect/lab/tree/shared/Live%20Figures%20Brain%20Responsiveness/model_gaba/live_figure.ipynb |
| Figure 3 panel B | https://t.ly/hHSIO | https://lab.jsc.ebrains.eu/hub/user-redirect/lab/tree/shared/Live%20Figures%20Brain%20Responsiveness/model_microcircuit/live_figure_microcircuit.ipynb |
| Figure 3 panel E | https://t.ly/NAnTy | https://lab.jsc.ebrains.eu/hub/user-redirect/lab/tree/shared/Live%20Figures%20Brain%20Responsiveness/model_mesocircuit/live_paper_mesocircuit_plotting.ipynb |
| Figure 4 panel C | https://t.ly/G-zOc | https://lab.jsc.ebrains.eu/hub/user-redirect/lab/tree/shared/Live%20Figures%20Brain%20Responsiveness/model_roques/mean-field/live_figure.ipynb |
| Figure 6 panel B | https://t.ly/n2OT1 | https://lab.jsc.ebrains.eu/hub/user-redirect/lab/tree/shared/Live%20Figures%20Brain%20Responsiveness/model_exploration/model_exploration_brainstates.ipynb |
| Figure 6 panel F | https://t.ly/evO0e | https://lab.jsc.ebrains.eu/hub/user-redirect/lab/tree/shared/Live%20Figures%20Brain%20Responsiveness/model_adex/model_adex.ipynb |
| Figure 7 panel A | https://t.ly/NMLWS | https://lab.jsc.ebrains.eu/hub/user-redirect/lab/tree/shared/Live%20Figures%20Brain%20Responsiveness/figure_multispecies/live_figure_unimi_25-v2.ipynb |
| Figure 8 all panels | https://t.ly/dI7qK | https://lab.jsc.ebrains.eu/hub/user-redirect/lab/tree/shared/Live%20Figures%20Brain%20Responsiveness/figure_wholebrain/00_figure_7.ipynb |

*Table 1: Links to the "live panels" of the figures of the article. The different panels indicated correspond to a model or analysis that can be run online in EBRAINS. The EBRAINS Collab where Live Figures are hosted is: https://wiki.ebrains.eu/bin/view/Collabs/live-figures-brain-responsiveness/ in the Drive tab. An EBRAINS account is needed to run the notebooks at lab.ebrains.eu.*

## Abstract

The objective of this paper is to review physiological and computational aspects of the responsiveness of the cerebral cortex to stimulation, and how responsiveness depends on the state of the system. This correspondence between brain state and brain responsiveness (state-dependent responses) is outlined at different scales from the cellular and circuit level, to the mesoscale and macroscale level. At each scale, we review how quantitative methods can be used to characterize network states based on brain responses, such as the Perturbational Complexity Index (PCI). This description will compare data and models, systematically and at multiple scales, with a focus on the mechanisms that explain how brain responses depend on brain states.

# Introduction

Spontaneous brain activity emerges from the interplay of neuronal intrinsic electrophysiological properties and their synaptic interactions, all under the dynamic influence of neuromodulatory systems, as discussed in the accompanying paper (Sanchez-Vives et al., 2025). Foundational intracellular recordings demonstrated that individual thalamic and cortical neurons have intrinsic membrane properties that drive ongoing activity even in the absence of inputs i.e., pacemaking properties in thalamic neurons due to the interplay between low threshold calcium current and h-current (Llinás, 1988; Llinás & Jahnsen, 1982). Subsequent *in vivo* studies revealed that synchronized synaptic barrages between recurrently connected neurons generate large-scale cortical-thalamocortical rhythms (Steriade et al., 1993), which are shaped by cholinergic, noradrenergic, and other modulatory pathways (Dan et al 2012; McCormick, 1992; Steriade et al 1993). More generally, spontaneous activity is strongly correlated with the brain functional state (Raichle et al., 2001; Steriade, 2000) and behavior (Stringer et al., 2019). Importantly, spontaneous activity can bias how external inputs are processed, such that internal dynamics determine the response to external inputs (Arieli et al., 1996). The relationship between spontaneous activity and brain responsiveness is the focus of this review article.

Much of the evidence on the relationship between evoked and spontaneous activity comes from the study of waking and sleeping states (reviewed in (Steriade et al., 2001); Figure 1). In particular, classic studies in visual (Arieli et al., 1996; Funke and Eysel, 1992; Li et al., 1999; Livingstone and Hubel, 1981; Tsodyks et al., 1999; Worgotter et al., 1998), somatosensory (Morrow and Casey, 1992), auditory (Edeline et al., 2001; Kisley and Gerstein, 1999; Miller and Schreiner, 2000), and olfactory systems (Murakami et al., 2005) have shown that high-amplitude slow-wave activity in electroencephalogram (EEG) measurements is generally associated with rather preserved local neural responsiveness (Figure 1A,B). Interestingly, Figure 1A illustrates the spectrograms during auditory stimuli in response to music during wakefulness (left) and non-rapid eye movement (NREM) sleep, showing high similarity in the high-gamma band between both states (Figure 1B) (Hayat et al., 2022). Sleep only moderately attenuated response magnitudes, mainly affecting late responses beyond early auditory cortex. These findings are consistent with intracellular recordings observed by Steriade *et al.* (Steriade et al., 2001), where neurons displayed similar excitability—as spikes evoked by intracellular pulses—in the awake state and slow wave sleep (SWS). In their observations, larger hyperpolarization of the membrane in SWS was counterbalanced by the increased conductance in wakefulness, due to membrane depolarization and synaptic bombardment (Figure 1C,D). Therefore, local responsiveness can be similar in SWS and in wakefulness.

However, the global brain response including interareal activations (Ferrarelli et al., 2010; Massimini et al., 2005) and feedback modulation (Hayat et al., 2022) are significantly reduced both in SWS (Figure 1E) and anesthesia (Figure 1F) with respect to wakefulness. High-density EEG recordings during transcranial magnetic stimulation (TMS) revealed that during quiet wakefulness, the initial local response was stronger in SWS than in wakefulness, but it was rapidly extinguished and did not propagate beyond the stimulation site. Conversely, during wakefulness, the local response was followed by a sequence of waves that propagated to both nearby and distant connected cortical areas (Figure 1E; Massimini et al 2005). Thus, the fading of consciousness during certain stages of sleep may be related to a breakdown in cortical effective connectivity, rather than to a decrease in the local response. The same has been observed under anesthesia, which also induces slow waves synchronization (Figure 1F; (Ferrarelli et al., 2010)). Furthermore, stimulus-evoked timing precision and neural selectivity are disrupted in SWS and anesthesia, where highly synchronous patterns of rhythmic activity dominate cellular membrane potential (Contreras and Steriade, 1997), making the network unreliable and less responsive to inputs.

In wakefulness and rapid-eye-movement (REM) sleep, neuronal activity is desynchronized and irregular (Harris and Thiele, 2011; Renart et al., 2010). In these states, the membrane potential is depolarized and close to the spike threshold, which explains why neurons respond more reliably and with less response variability to inputs compared with slow-wave sleep. However, even in wakefulness, variations in the spatiotemporal patterns of spontaneous activity can strongly influence information processing, and conversely, sensory inputs can alter ongoing activity. Such interplay between intrinsically generated activity and its modulation by external inputs is a central aspect of the mechanisms by which the brain processes external inputs. New methods for characterizing the complexity of network dynamics and their response patterns have emerged, particularly recently (Krohn et al., 2023; Wolf et al., 2018), and are presented here at various scales. Here, complexity is associated with the set of accessible states of a system (Parisi, 2006). In the present context, this notion can be linked to the diversity of slow-wave propagation modes and to the richness (i.e., the entropy) of perturbation-evoked responses in brain activity. It is also important to recognize that although pioneering intracellular studies *in vivo* have explored the cellular mechanisms of synchronized oscillations during sleep and anesthesia and have attempted to explain why neuronal responsiveness is different during these states (Reig et al., 2015; Steriade, 2000), much less is known about the dynamics of wakefulness. The main reason for this is that it is very difficult to perform stable intracellular recordings in awake animals. This caveat has been partially overcome in recent years by the advent of new recording techniques such as multiple extracellular recordings, various imaging techniques based on calcium or voltage indicators, or large-scale imaging using global brain signals (Afrashteh et al., 2021; Brier et al., 2019; Celotto et al., 2020; Filipchuk et al., 2022; Goltstein et al., 2015; Mohajerani et al., 2013; Montagni et al., 2024; Resta et al., 2022; Tort-Colet et al., 2023).

Finally, computational models have also shown that stochastic-like or highly chaotic network states can exhibit increased responsiveness. Early studies have shown that chaotic neural networks exhibit enhanced information processing capabilities (Arnold et al., 2013; Bertschinger et al., 2004; Destexhe, 1994; Destexhe and Contreras, 2006; Vreeswijk and Sompolinsky, 1996). These studies have highlighted not only the fact that neurons are depolarized, but also that they exhibit strong fluctuations in membrane potential, and it is the presence of these fluctuations that places neurons in a highly responsive mode (Ho and Destexhe, 2000; Kuhn et al., 2004) with precise spike timing (Nowak et al., 1997).

In this review, we provide an overview of some of the methods used to characterize brain complexity and show how they can be used to characterize brain responsiveness at different levels by mixing experiments and models, from circuits to mesoscales to the whole brain. Particular emphasis is placed on perturbational measures of complexity, which quantify the deterministic causal interactions revealed by a controlled perturbation of the system. Unlike observational measures, which are computed from spontaneous activity and capture statistical dependencies within neural signals, perturbational approaches provide a framework to investigate how causal complexity changes across spatial scales (Sarasso et al 2021.). For this reason, throughout this Review we focus primarily on the Perturbational Complexity Index (PCI) and related perturbational complexity measures (see box 2 for details) across scales. We define the scales as follows: “microscale” refers to the level of single cells up to local networks of neurons, “mesoscale” from several local networks (or columns) to a single brain area, and “macroscale” from several brain areas to the whole brain. For each level, selected figure panels will be linked to interactive components that can be executed and reproduced online via the EBRAINS neuroscience platform. In addition, some of the underlying data and models are publicly available.

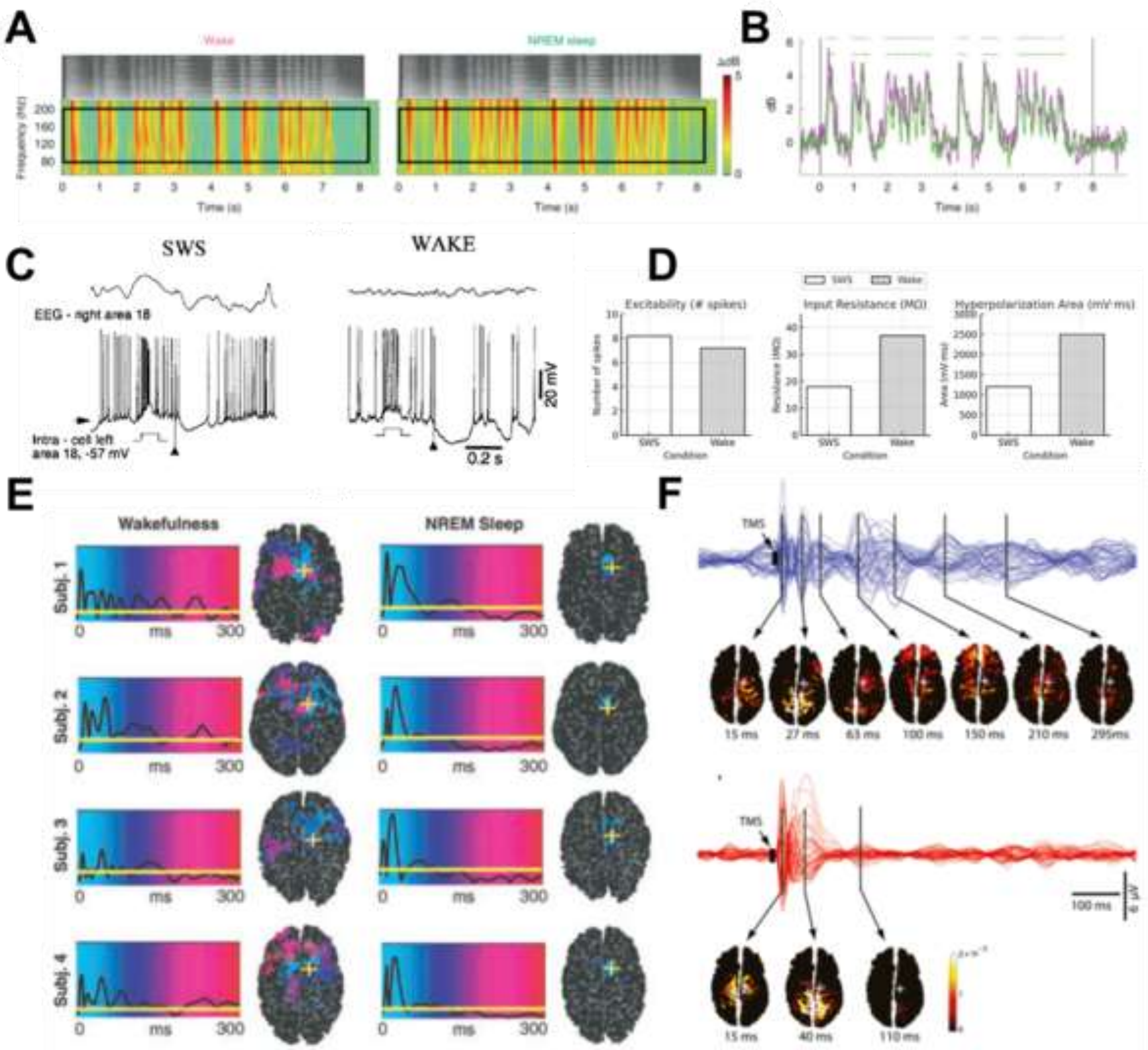

***Figure 1. Dissociation of local versus global cortical responsiveness across wakefulness, NREM sleep, and anesthesia.*** *Panels illustrate that although local neuronal and high-frequency responses remain relatively preserved between wake and sleep or under anesthesia, large-scale cortical propagation is selectively disrupted during both NREM sleep and anesthetic-induced unconsciousness.* ***(A)*** *Time–frequency spectrograms of human auditory cortex (80–200 Hz gamma band) in the same stimulus window during wakefulness (left) and NREM sleep (right), showing robust local gamma responses in both states. Adapted from (Hayat et al., 2022).* ***(B)*** *Trial-averaged gamma-band amplitude (dB) over time (solid lines) and baseline confidence intervals (dashed lines; wake: magenta, sleep: green), demonstrating preserved local gain but attenuated feedback modulation during sleep. Adapted from (Hayat et al., 2022).* ***(C)*** *Intracellular recordings of a cat neocortical neuron during slow-wave sleep (SWS; left) and quiet wakefulness (WAKE; right). Top traces, surface EEG; bottom traces, membrane potential showing stimulus-triggered down-state (arrow) followed by recovery and firing. The stimulus (triangle) corresponds to the electrical shock to the homotopic contralateral area 18. Adapted from (Steriade et al., 2001).* ***(D)*** *Summary of local excitability measures in SWS versus wake: number of spikes evoked by a standardized current pulse (left), somatic input resistance (middle), and hyperpolarization area (right). Adapted from (Steriade et al., 2001)*. ***(E)*** *Transcranial magnetic stimulation (TMS)–evoked potentials and source-reconstructed cortical maps in four human subjects during wakefulness (left) and NREM sleep (right), showing widespread propagation when awake and collapse to a local response during sleep. The colour code corresponds to the latency of activation (light blue, 0 ms; red, 300 ms). The Y-axes is the global mean field power. Adapted from (Massimini et al., 2005).* ***(F)*** *TMS-EEG responses in human subjects in awake (blue) vs midazolam sedation (red): top, individual single-trial waveforms superimposed*

*(blue/red) with average (black); bottom, cortical source maps at representative latencies, revealing preserved local activation but loss of global spread. Adapted from (Ferrarelli et al., 2010).*

## Responsiveness at the cellular and local circuit levels

We first examine the responsiveness of local networks of cortical neurons, contrasting models and experiments. As discussed in the previous section at the whole-brain level, brain state also determines the stimulus-evoked response in local circuits. One way to study these circuits is to use cortical slices, in which the circuitry—preserving columns and layers—retains much synaptic connectivity while being disconnected from the rest of the brain. Interestingly, this local circuitry is sufficient to generate spontaneous slow oscillations that closely resemble those observed during slow-wave sleep and anesthesia, consisting of Up (active) and Down (silent) periods (Sanchez-Vives et al., 2017; Sanchez-Vives and McCormick, 2000). Adding neurotransmitters that promote arousal such as acetylcholine and norepinephrine to the slices successfully desynchronizes the slow waves and induces an awake-like pattern (Barbero-Castillo et al., 2021; Constantinople and Bruno, 2011; D'Andola et al., 2018) (Figure 2A). This approach has enabled investigation of the spatiotemporal response to electrical stimulation of the network in different dynamical regimes, such as slow waves versus awake-like activity (Figure 2A). These regimes represent extremes of the multidimensional space in which cortical activity resides (Harris and Thiele, 2011). As described above for the whole brain (Ferrarelli et al., 2010; Massimini et al., 2005), electrical stimulation of the local cortical network in the slow-wave regime evokes a large-amplitude, synchronized response that soon collapses, whereas stimulation in the awake-like state produces a response that reverberates throughout the network and exhibits higher spatiotemporal complexity than in slow waves (D'Andola et al., 2018).

In slices, the physiological processes by which the brain switches between dynamical states can be induced by various means including pharmacological manipulation (Figure 2B). At one extreme of the spectrum, slow oscillations represent a low-complexity state, as assessed by sPCI (slice Perturbational Complexity Index; D'Andola et al., 2018). We use sPCI to refer to the PCI measure adapted for application to cortical brain slices (see Box 2). It relies on the same core algorithm as PCI, namely the Lempel–Ziv complexity of the spatiotemporal pattern of significant responses (Casali et al., 2013) but differs in the preprocessing steps required for slice recordings. A more desynchronized state—achieved in the presence of norepinephrine and acetylcholine—represents the opposite extreme and has higher complexity. The key role in increasing complexity may be associated with the heightened excitability caused by norepinephrine and cholinergic agonists. However, the scenario is not so simple, since pharmacological enhancement of cortical excitability by other means (e.g., with kainate) does not significantly modify cortical complexity (Figure 2C). Furthermore, extremely enhanced excitability, as in epileptic discharges, is associated with lower complexity (Escalona-Morán et al., 2010). Therefore, sleep-like bistability; that is, the tendency to fall into a silent Down state, seems to play an important role in the breakdown of causality and in reducing perturbational complexity (D'Andola et al., 2018).

Both cholinergic and adrenergic systems enhance cortical complexity, an effect also attributable to their modulation of specific ionic currents such as the M-current and h-current (Dalla Porta et al., 2025, 2023)—that play important roles in the modulation of Up and Down states. Cortical processing is as well highly dependent on the co-occurrence of excitation and inhibition. The roles of slow and fast inhibition—mediated by $GABA_A$ and $GABA_B$ receptors, respectively—have been studied in cortical slices departing from two distinct states: synchronized (slow oscillations) and desynchronized (awake-like activity) (Barbero-Castillo et al., 2021). Independently of dynamical state, fast inhibition affects cortical complexity: its blockade through gabazine (GBZ) results in lower sPCI values (Figure 2D,E). Conversely, slow inhibition impacts complexity only during synchronized states, where its blockade also reduces sPCI. Altogether, balanced inhibition in cortical activity is crucial for providing richness in the emergent patterns, contributing to the complexity of causal interactions.

The relationship between excitatory and inhibitory balance has also been explored in a data-driven *in silico* model (Figure 2F). The perturbational complexity observed in *in vitro* models has been reproduced *in silico* (Figure 2G). The computational model enables exploration of areas in a parameter space not accessible experimentally. By varying the strength of fast GABAA-mediated inhibition around its physiological reference value, the model shows that perturbational complexity is maximal within an intermediate E/I-balance window. Reducing inhibition, as in disinhibition, and increasing inhibition beyond this reference both lower sPCI, although for different reasons: disinhibition favors excessive synchrony, whereas excessive inhibition prevents propagation of the perturbation. This balance is found around physiological values (Figure 2H).

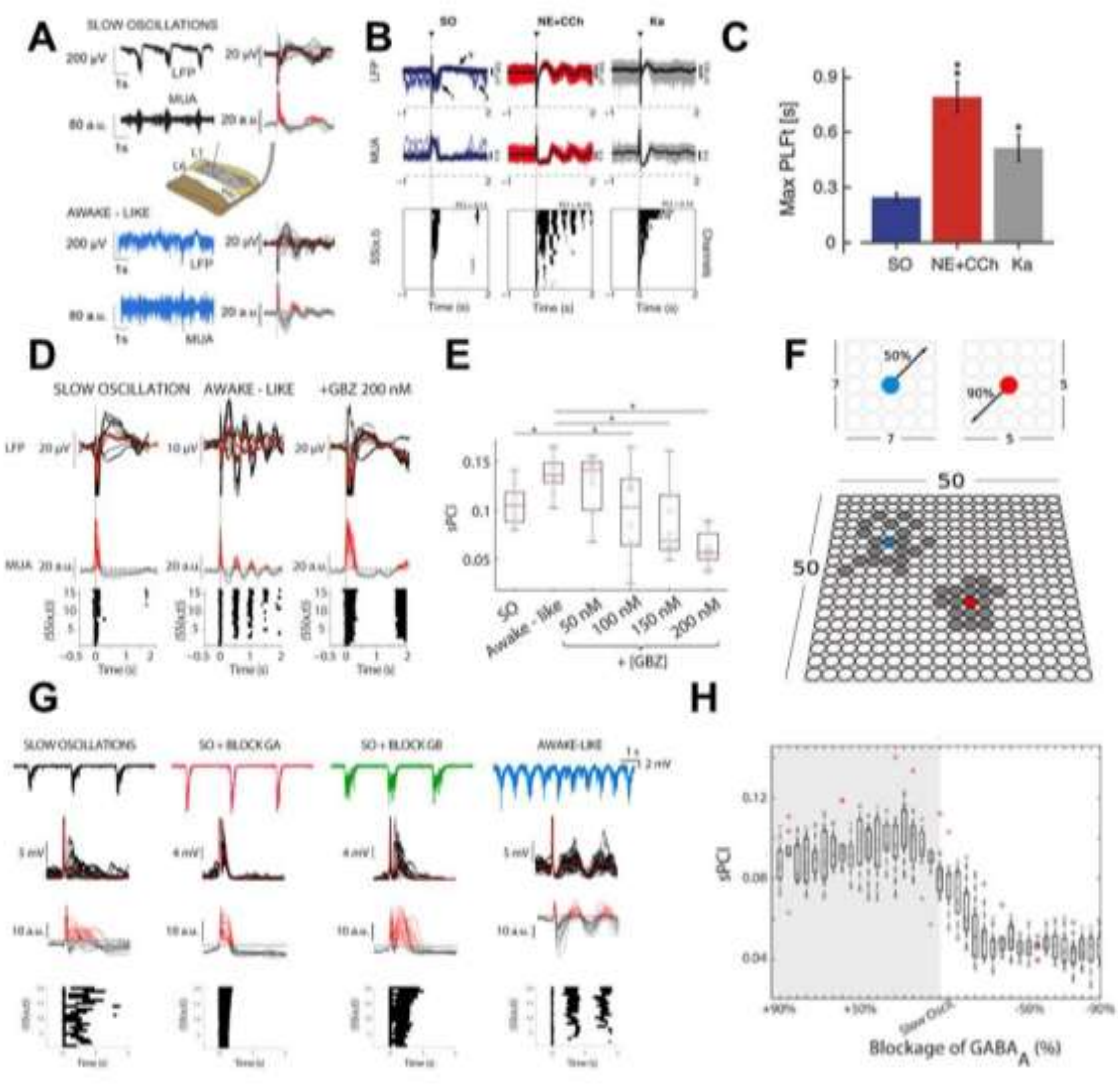

***Figure 2. Responsiveness in local circuits across different conditions: in vitro experiments and models. (A)*** *Neocortical slices activity recorded with 16-channel multielectrode array (MEA). Single pulses of electrical stimulation were applied to the infragranular layers. Raw local field potential (LFP) traces during the regime of slow*

*oscillations (SO) at the top and multiunit activity (MUA) at the bottom. The same is displayed below (blue) for desynchronized activity, awake-like, in the presence of neurotransmitters. On the left, recordings of 5s of spontaneous SO (top), and NE+CCh (bottom). On the right, single traces and averaged (red) LFP (top) and MUA (bottom) responses to electrical stimulation representing spatiotemporal responses across the slice. Adapted from (Barbero-Castillo et al., 2021). (**B)** Responsiveness of cortical slices under three different conditions: SO, NE+CCh, and Ka. The responses to electrical stimulation recorded by a single channel are shown both in the LFP and MUA in the first two rows, respectively. The black lines indicate the average response for each condition. Below, the corresponding binary matrix of significant responses to the stimulation [SS(x,t)]. (**C)** Histogram showing the population values of sPCI averaged across slices (n=14). Adapted from (D'Andola et al., 2018). (**D)** Averaged LFP (top) and MUA (bottom) responses to electrical stimulation during spontaneous SO (left), desynchronized activity (middle), and with bath application of 200nM of GBZ (right). Binary matrices of significant sources of activity [SS(x,t)] following electrical stimulation delivered to neocortical slices (bottom). **(E)** Population values of sPCI shown as boxplot for increasing GBZ concentrations (50, 100, 150, and 200nM GBZ *p=<0.05). **(F)** Model of a cortical slice consisting of pyramidal (blue) and inhibitory (red) neurons arranged in a 50 × 50 square lattice. (**G)** Responsiveness of the in silico model reproduces what is observed experimentally during (from left to right) SO, SO + blocking $GABA_A$, SO + blocking $GABA_B$, and desynchronized activity. Single spontaneous sLFP (top), averaged sLFP (middle top), MUA (middle bottom), and binary matrices of significant sources of activity [SS(x,t)] (bottom) are shown for each condition. (**H)** Population sPCI computed on the response generated by the model for inhibited and disinhibited cortical networks. The gray area represents the model predictions. Adapted from (Barbero-Castillo et al., 2021).*

Similar results have recently been found by combining *in vivo* LFP recordings from layers of a single cortical area, with electrocorticography (ECoG) from several cortical regions in mice (Hönigsperger et al., 2024). Only during wakefulness was the response to the stimulation long-lasting and propagated within the cortical column to other brain regions. Interestingly, similar levels of spatiotemporal complexity were found at both local and global scales (Hönigsperger et al., 2024). Conversely, during slow-wave activity, the response was only brief and spatially localized (Hönigsperger et al., 2024). In the slow-wave regime, stimulation also triggered a silent, highly synchronized period that generated a refractory window for subsequent reverberations (Camassa et al., 2022).

Computational models investigated canonical circuits of cortical assemblies composed of excitatory neurons (regular-spiking (RS)) and inhibitory neurons (fast-spiking (FS)) with an adaptation mechanism of the firing rate, like short-term synaptic depression (STD) or potassium currents (voltage-dependent or calcium/sodium-mediated spike-frequency adaptation (SFA)), both of which are activity-dependent. A model integrating these basic mechanisms captures much of the richness of the dynamical phases observed across brain states (di Volo et al., 2019; Mattia and Sanchez-Vives, 2012; Torao-Angosto et al., 2021; Zerlaut et al., 2018).

The full spectrum of such dynamical phases can be exposed by modulating two specific features of the microscopic circuit: (1) the strength of the adaptation of the firing rate in the excitatory neurons, and (2) the excitation level of the same subpopulation of cells (Gigante et al., 2007; Latham et al., 2000; Mattia and Sanchez-Vives, 2012). On the one hand, a stronger excitation can be associated with an increase in spike rate from the input of excitatory neurons of other brain areas. On the other hand, a modulation of the adaptation strength changes the amount of activity-dependent hyperpolarization, dampening the firing rate of the cortical circuit under an active high-firing state (Figure 3A). In the excitation-adaptation plane, the bifurcation diagram of Figure 3B can be worked out in simulation and under mean-field approximation. The network of integrate-and-fire (IF) neurons can be trapped in an asynchronous-irregular (AI) state at low and high firing rate (top-left and bottom-right phase, respectively) roughly modeling the burst-suppression (Amzica, 2015; Ching et al., 2012; Lewis et al., 2013) and the awake brain states, respectively. At relatively high levels of both excitation and adaptation (Figure 3B, top-right), the circuits spontaneously express "slow-fast relaxation" oscillations (Latham et al., 2000; Mattia and Sanchez-Vives, 2012) modeling the slow periodic alternation of high-firing (Up) and almost-quiescent (Down) states typically observed under NREM sleep and the unresponsive brain state induced by general anesthesia.

Close to transition points separating the slow oscillation (SO) phase from the awake-like phase, spontaneous activity is likely desynchronized, although a richer dynamical diversity can be expressed by local circuits. To elicit such additional modes of activity, a focal stimulation sparse in time (Figure 3C) can be instrumental in describing the global brain state with higher resolution (Cattani et al., 2023). This is because the response to external perturbations depends on the time course of past activity, which in turn is shaped by the adaptation level (Figure 3A-C). Although the spontaneous activity in phases (1) and (2) shown in Figure 3B,C appear to be similarly desynchronized, once perturbed they display a qualitatively different behavior. More specifically, when adaptation level is higher (2), stimulation elicits the occurrence of synchronized off periods.

Such state-dependent responsiveness is not only related to the global brain state but also to the microscopic level of adaptation which increases when the firing rate of the circuit in the recent past is persistently high and hence fatigued. As a result, during the SO brain phase the same stimulation can elicit an Up state if delivered at the end of a Down state, while the Up state will be terminated if the stimulation is received by an active (and fatigued) circuit (Curto et al., 2009; Linaro et al., 2011). What was predicted by this theoretical framework has been found experimentally in cortical slices spontaneously expressing both the synchronized sleep-like and the desynchronized awake-like state (Figure 2A,B; (Barbero-Castillo et al., 2021)).

The responsiveness in different global states in networks of spiking (i.e., integrate-and-fire) neurons can be further modulated by the presence of other microscopic features which in turn are differently expressed depending on the level of activity. A paradigmatic example is in Figure 3D**,** where the response to a Gaussian-distributed excitatory input is shown in two different desynchronized (AI) states in a network of AdEx neurons (di Volo et al., 2019). The spike rasters of units show that the same input evokes a different response in the two network states. This is particularly visible when computing the instantaneous firing rate of the excitatory neurons (bottom, noisy curves). The continuous curves show that the AdEx mean-field model can capture the response in the two different network states, and thus accounts for state-dependent responsiveness.

An example describing local cortical circuits with more detailed connectivity is the spatially structured, multi-layer cortical network of integrate-and-fire neurons illustrated in Figure 3E (Senk et al., 2024). Here, the cortical response to a local thalamic input during a desynchronized-like brain state can indeed uncover the existence of a nearby metastable Down state which may propagate away from the site of thalamic activation, sustained by distance-dependent connectivity.

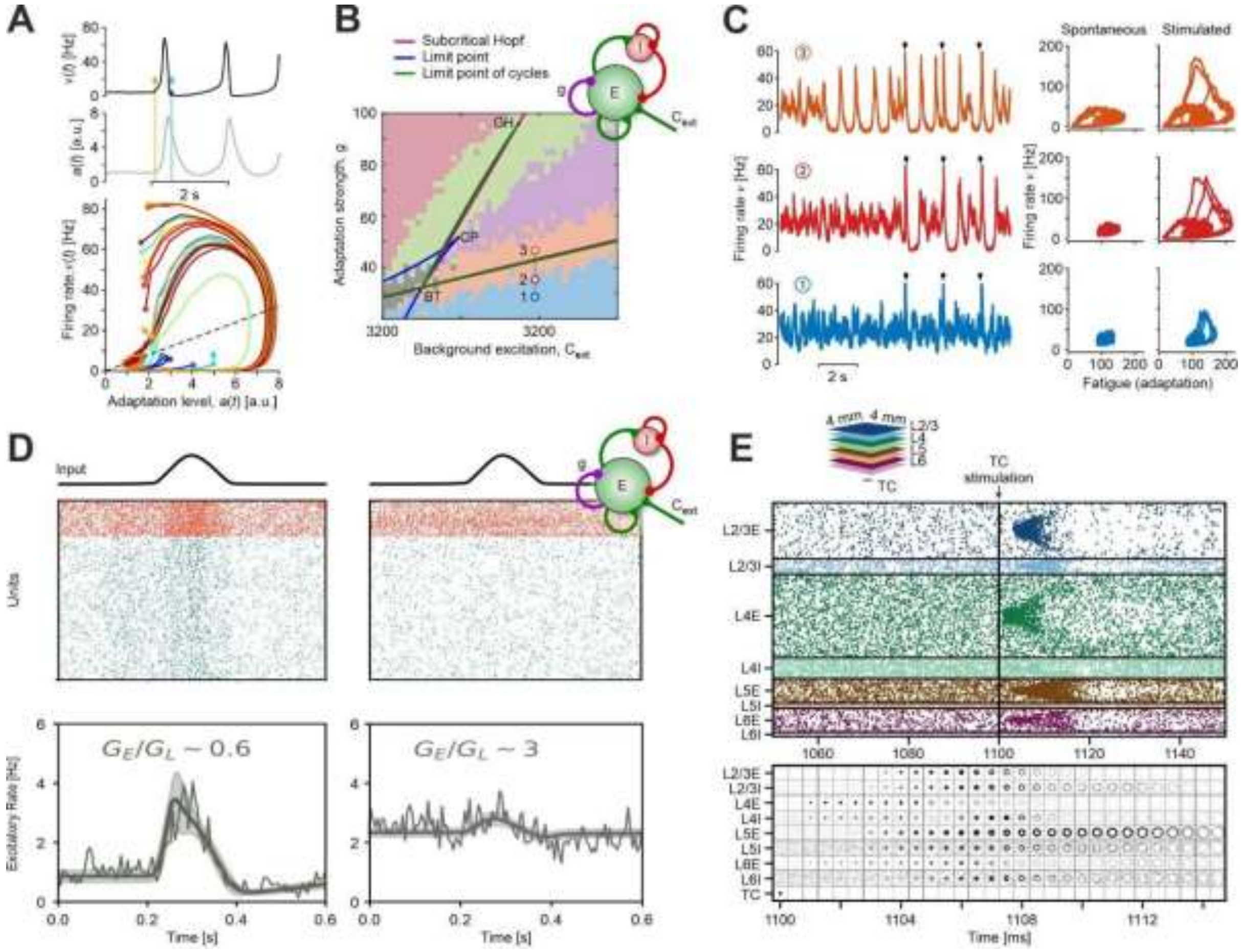

***Figure 3. Responsiveness across brain states at the circuit level.*** *(**A**) Top, firing rate ν(t) from a network of 20,000 excitatory integrate-and-fire (IF, spiking) neurons. Arrows and dashed lines, time of perturbation induced by a sudden change in the synaptic current. Middle, Average adaptation level a(t). Bottom, post-perturbation trajectories in the (a, ν) plane showing the state-dependent nature of the responses. Adapted from (Linaro et al., 2011). (**B**) Bifurcation diagram showing the different activity regimes displayed by network simulations of about 9,000 spiking (excitatory and inhibitory leaky IF) neurons as a function of the firing rate adaptation and the number of incoming excitatory Poisson processes from other (external) areas. Solid lines show the transitions predicted by the theory while the colors indicate the different dynamical regimes the "finite size" spiking neural network exhibits in simulations. (**C**) Left, the firing rate of excitatory spiking neurons is shown for three simulations associated with different positions on the phase diagram of panel A. The black arrow indicates the moments in which the network receives a precise excitatory stimulus. Right, the same trajectories are shown on the 'Firing Rate' vs. 'Adaptation Current' plane. B and C adapted from (Cattani et al., 2023). (**D**) Left and right columns show two different asynchronous irregular (AI) states in a network of 10,000 adaptive exponential (AdEx) integrate-and-fire neurons (8,000 excitatory and 2,000 inhibitory) in two different asynchronous-irregular network states (corresponding to different parameter values). The response to a Gaussian-distributed excitatory input (top) is indicated in the raster of units (middle). The instantaneous excitatory firing rate is shown in the bottom (noisy curves), together with the AdEx mean-field model (continuous curves). Adapted from (di Volo et al., 2019). (**E**) Layered spiking network model covering 4 × 4 mm² at biologically realistic neuron density. Activity is evoked at t = 700 ms by a spatially and temporally confined thalamic pulse. Evoked activity propagates through the cortical network as displayed in the spike raster plot and sequence of snapshots of spatiotemporally binned firing rates. Adapted from (Senk et al., 2024).*

## Responsiveness at the mesoscale level

We now examine responsiveness at a larger scale, spanning millimeter distances, which is often referred to as the mesoscale. We first show that responses to sensory inputs can occur as traveling waves, in the awake animal. This was shown for the primary visual cortex (V1) of an awake monkey during a fixation task (Muller et al., 2014). Traveling waves can occur either spontaneously (Figure 4A) or in response to a visual input (Figure 4B). While the spontaneous traveling waves occurred randomly, with random initiating sites in the cortex, and random propagation directions (Figure 4A), the evoked traveling waves were more structured (Figure 4B). Remarkably, nearly all visual inputs evoked a traveling wave, and these different waves were always similar in initiation site and direction of propagation (although with some variability between trials, especially in latency). These waves were modeled by the AdEx mean-field models (Zerlaut et al., 2018), as illustrated in Figure 4C. An example of a traveling wave evoked by a single visual input (Figure 4C, top rows) is contrasted with more complex patterns of traveling waves evoked by multiple inputs (Figure 4C, bottom rows). These two cases were modeled by a 2D array of mean-field models where the connectivity was adjusted to the propagation velocity and extent of the propagating waves (Zerlaut et al., 2018). It must be noted that this model also accounts for the suppressive interactions between traveling waves (Chemla et al., 2019), which are also apparent in the activity evoked by multiple visual inputs (Figure 4C, bottom rows).

Spontaneous slow waves can also display propagating patterns, as shown in anesthetized mice (Huang et al., 2010; Mohajerani et al., 2010; Pazienti et al., 2022; Stroh et al., 2013). Under relatively deep anesthesia, spiral waves occur more frequently than planar waves. The opposite unbalance is observed as a lighter level of anesthesia is administered (Huang et al., 2010). These spiral waves in cortex are thought to be generated by a different spreading of the lateral excitation in the vicinity of the borders of a cortical area like the primary visual cortex (Ermentrout and Kleinfeld, 2001; Huang et al., 2010; Schiff et al., 2007). In the absence of such heterogeneity, spiral waves are thought to be unlikely. This expectation is confirmed by inspecting slow-wave activity across the whole cortical hemispheres (Greenberg et al., 2018; Liang et al., 2023) and higher-order cortical areas (Dasilva et al., 2021; Pazienti et al., 2022) at different levels of anesthesia. In these cases, only planar waves have been found. This is predicted by models of cortical areas where spatially organized networks of excitatory and inhibitory spiking neurons with adaptation quantitatively mimic slow-wave activity observed in mice (Figure 4D-left; Pazienti et al., 2022). Interestingly, these *in silico* cortical areas predict that in the absence of peculiar excitability of the borders, focal stimulation can in principle make apparent the state-dependency of slow-wave patterns (Figure 4D-right, (Galluzzi et al., 2025)).

***Figure 4. Cortical responses occurring as traveling waves.*** *(**A**) Spontaneous traveling waves occurring in the awake monkey primary visual cortex, imaged by VSD. Each panel shows a phase latency map for three examples of spontaneous traveling waves. (B) Traveling waves evoked by visual inputs. Three examples of the same visual input and the traveling wave evoked. (C) Model of traveling waves using a 2D array of AdEx mean-field units. The top row shows snapshots of the simulated VSD signal for a single visual input evoking a traveling wave. The bottom row shows the response evoked by more complex inputs in the same conditions. (D) Cortical area modeled as a lattice of cortical assemblies. Changing both adaptation and background excitation, different slow-wave activities are induced which are reminiscent of in vivo observations under deep and light anesthesia (Pazienti et al., 2022). A focal stimulation randomly involving one cortical assembly (red dots) elicits slow activation waves with a probability depending on the past activity of the cortical field. The stimulation likely elicits a spiral wave (top right snapshots) under deep-like anesthesia state, whilst a planar wave (bottom right snapshots) is evoked under the light-like anesthesia state. For both C and D, the population models were composed of excitatory and inhibitory IF neurons incorporating spike-frequency adaptation and being excited by an external source of excitatory neurons. (A,B: adapted from (Muller et al., 2014); C: adapted from (Zerlaut et al., 2018); D: adapted from (Pazienti et al., 2022) and (Galluzzi et al., 2025).*

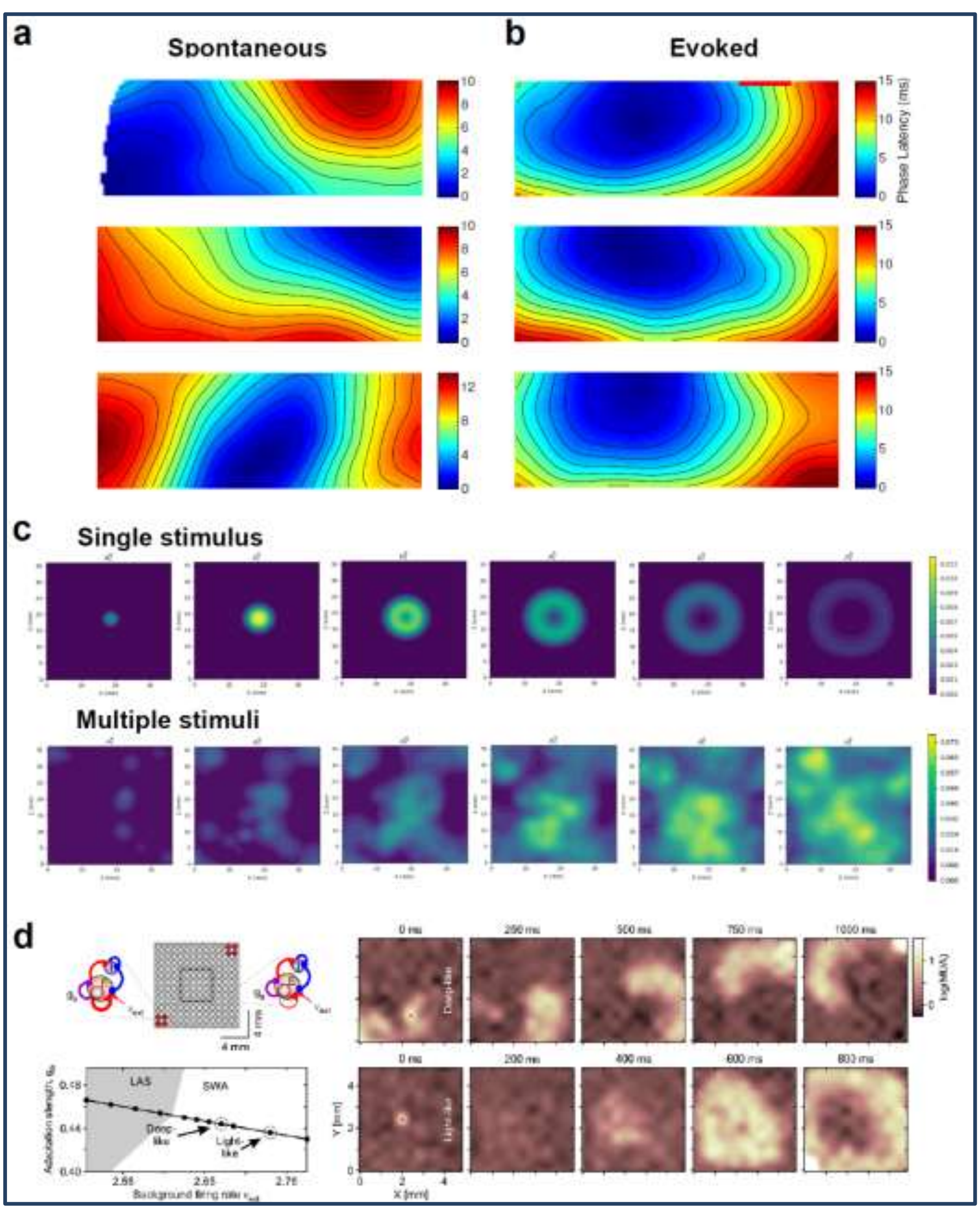

Besides propagating patterns, cortical responses can also occur in a more distributed fashion. At the level of a single cortical area, responses to direct electrical stimulation largely depend on the brain state. We illustrate here this statement by considering, in non-human primate and in implanted patients with epilepsy, three cases of brain state transitions induced by (1) anesthesia (Dwarakanath et al., 2025); (2) epilepsy (David et al., 2008; Russo et al., 2023); (3) wake-sleep (Pigorini et al., 2015). Beyond clear transitions, we have also gathered evidence that local and remote responses to direct electrical stimulations are variable, even with no obvious behavioral state changes (patient at rest) (Jedynak et al., 2023).

Cortical responses were investigated using Utah-arrays of microelectrodes in the monkey cortex. Because the evoked spiking response to microstimulation displayed complex, and putatively non-linear transfers, analysis of the spread of propagation was restricted to local field potentials (LFPs) (Figure 4A-D). While in the ventrolateral prefrontal cortex (PFC), the modulation strength at distal populations was equivalent to that at proximal populations in wakefulness (Figure 4A,C), it decayed with distance in the posterior parietal cortex (PPC) (Figure 4B,D). However, the modulation decayed sharply over distance as a function of the depth of the anesthesia in both the PFC and PPC, thus pointing to distinct roles of the two nodes of the fronto-parietal loop in conscious wakefulness (Dwarakanath et al., 2025).

Single-area measurements can also be obtained in human subjects using stereo-electroencephalographic (sEEG) evaluations for epilepsy surgery. Here, direct electrical stimulation has been used to induce seizures resembling spontaneous ones and thereby identifying the seizure onset zone (Kahane et al., 1993). In mesial-temporal lobe epilepsy, it was shown that a short-term plasticity can be induced by repeated stimulations at 1 Hz, which leads to the pre-ictal to ictal state transition (David, 2007; David et al., 2008; Russo et al., 2023) shown in Figure 5E. During this transition, (1) initially, the brain is moderately excitable and responds in a quasi-linear regime; (2) it becomes more excitable and the amplitude of responses to stimulation grows. This effect is shown in the top panel of Figure 5E**.** Finally, (3) the seizure starts and the area is no longer responsive to external inputs, as can be seen in the orange parts of the middle and bottom panels in Figure 5E.

Simple neural mass modeling (Jansen and Rit, 1995) can replicate this change of responsiveness by using the model in a regime close to a state transition, in the case presented here, marked by a saddle-node on invariant circle bifurcation (Grimbert and Faugeras, 2006; Touboul et al., 2011). This transition occurs between regimes III and IV marked in the model's bifurcation diagram (Grimbert and Faugeras, 2006); here adapted from (Jedynak et al., 2017)) shown in Figure 5F. The three insets show three responses of the model to external stimulation (pulse with amplitude 10 $s^{-1}$ and duration 200 ms delivered at time $t$=0), obtained in three different dynamical regimes. These responses correspond qualitatively to the three behaviors presented in Figure 5E and described above: (1) when the model operates in a quasi-linear regime II, it is moderately responsive to external input (bottom inset), (2) when the model operates in the excitable regime III, its response to stimulation is maximal (middle inset), (3) when the model operates in the limit cycle (regime IV), thus generating spike and wave like oscillations (i.e., seizure), it is not responsive to transient inputs (top inset). Recently, it was shown that the change of responsiveness of mesial-temporal lobe seizures may not be generalizable to neocortical seizures, for reasons that remains to be further elucidated (Russo et al., 2023).

In Pigorini et al. (Pigorini et al., 2015), the authors compared the cortico-cortical evoked potentials (CCEPs) recorded during wakefulness and NREM sleep by means of time–frequency analysis (Delorme & Makeig 2004) and Phase Locking Factor (PLF) (Lachaux et al., 1999) in eight patients with epilepsy implanted with stereo EEG electrodes for clinical evaluation. They observed that during wakefulness single pulse electrical stimulation (SPES) triggers a chain of sustained effects, as indicated by a phase-locked response that lasted for about 0.5 s. During NREM, the same initial activation induces a slow wave and a cortical OFF-period in its cortical targets after which the phase-locked response breaks-off, in spite of restored levels of cortical activity. Importantly, when SPES is delivered close to the recording site, the PLF is long lasting and remains significant until −500 ms during wakefulness even if the cortical neurons react to stimulation with a slow wave and a suppression of high frequency similar to NREM. This is possibly due to the local paraphysiological effects of intracranial electrical stimulation (Borchers et al., 2012). A plausible explanation for the persistence of deterministic effects induced by SPES following this local OFF-period is the feedback of phase-locked activity from the rest of the network during wakefulness. Coherently, the same early interruption of the phase-locked response to the stimulation was seen with general anesthesia, at the level of single cortical area in rats (Arena et al., 2021), and across layers from a single cortical region in mice, resulting in a local sequence of neuronal activations of reduced complexity, compared with wakefulness (Hönigsperger et al., 2024).

***Figure 5. Cortical mesoscale responsiveness at different brain state levels in animal and human data and simulations.*** *(A) State characteristics of mesoscale unit firing patterns in monkey PFC recorded using Utah-arrays. (B) Similar recordings in monkey PPC. In both areas, multi-unit spiking activity was monitored across the three levels of consciousness (quiet wakefulness—red, light anesthesia—green, deep anesthesia—blue). Black dots represent neuronal spikes. Sparsely-firing units during wakefulness were discarded, and this population was fixed for analysis during the anesthetic states. The red, green and blue rectangles denote the path of the Up state (P(UP) vs P(DOWN)), as estimated by a two-state Gaussian hidden Markov model. Progressively, the spiking activity of the network reorganizes into distinct periods of high and low activity, with UP and DOWN states (C) Field activity propagation within the PFC after electrical stimulation. (D) Field propagation in PPC. The red, green and blue curves represent the modulation in signal energy of the broadband (0.1-200Hz) LFPs over distance. The inset bar graphs show the total rate of change in signal energy modulation for proximal (first three spatial bins) and distal (last three spatial bins) populations. In the PFC, signal energy modulation is maintained across proximal and distal populations during wakefulness but decreases as a function of the depth of anesthesia. However, in the PPC, the signal energy modulation decreases with distance in all three states, with similar modulation strengths during wakefulness and deep anesthesia (adapted from (Dwarakanath et al., 2025). (E,F) Responsiveness to electrical stimulation increases during pre-ictal state and is inhibited during ictal state. (E) Experimental data. The top time trace (David, 2007) shows SEEG responses in the anterior hippocampus during 1Hz stimulation of the amygdala. The amplitude of the response grows until the occurrence of the seizure (around 24 s). The middle panel (Russo et al., 2023) shows response to stimulation in the medial temporal lobe. Stimulation time points are marked with vertical magenta dashed lines. The two bottom plot rows show this response averaged over all trials and overlaid for all recording contacts. (F) Theoretical framework. The bifurcation diagram of the Jansen & Rit model with three insets showing outputs of the model operating in three different dynamical regimes (marked with roman numbers) differing only in the value of the model's baseline input (abscissa of the main plot). In each case a stimulation pulse with amplitude 10 $[s^{-1}]$ and duration 200 ms was delivered at time t = 0. The black arrows link the value of the baseline input (60 $[s^{-1}]$, 110 $[s^{-1}]$ and 120 $[s^{-1}]$) with the corresponding generated time traces. Adapted from (Jedynak et al., 2017).*

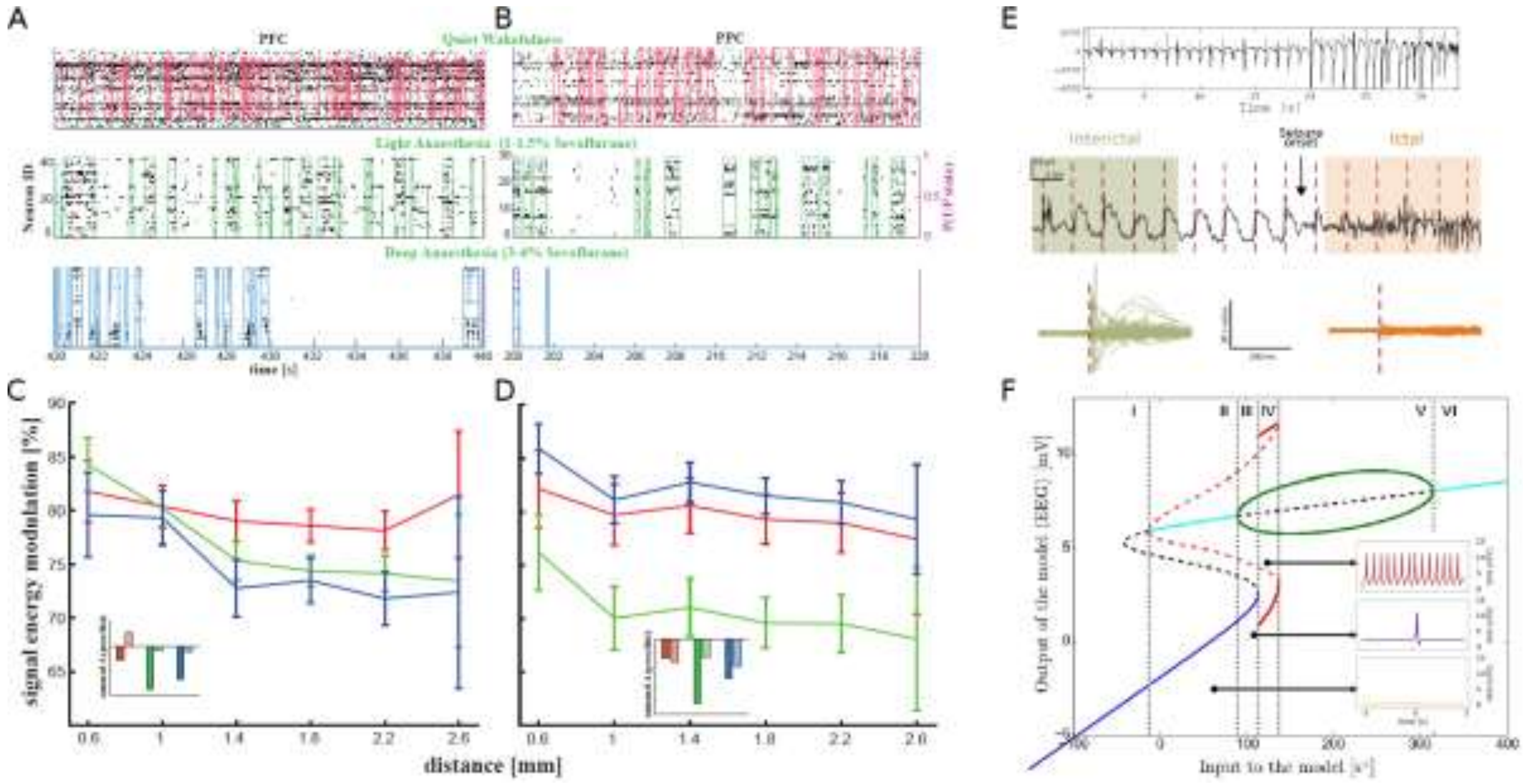

## Macroscale, from extended brain areas to the whole brain

Because complex cognitive processes like perception and learning rely on communication between multiple brain regions, investigating how sensory-evoked cortical activity patterns evolve over space and time across the entire brain is key to enhancing our understanding of these functions (Ferezou et al., 2007; Karimi Abadchi et al., 2020; Luczak et al., 2015; Mohajerani et al., 2013). Brain states—ranging from wakefulness to sleep, attention to distraction, or rest to active engagement—profoundly influence how neural circuits respond to external perturbations. Studying the brain-state dependence of exogenously evoked responses at the macroscale is crucial for understanding how the brain dynamically processes external stimuli under varying internal conditions.

Macroscale measurements, such as optical imaging, EEG, MEG, or fMRI, offer insights into how global functional connectivity and oscillatory activity modulate stimulus processing. Large-scale optical imaging techniques, combined with activity-dependent fluorescent dyes or genetically encoded indicators, have enabled the visualization of cortex-wide spatiotemporal activations in awake and anesthetized subjects (Cardin et al., 2020; Celotto et al., 2020; Gutzen et al., 2024; Montagni et al., 2018; Ren and Komiyama, 2021). With these high-resolution tools, it has been possible to analyze the brain-state dependence of stimulated responses distributed across the entire dorsal cortex (Liang et al., 2023; Montagni et al., 2024; Rosenthal et al., 2021; Song et al., 2018). These studies demonstrated that the spatiotemporal pattern of evoked activity propagation strongly depends on the brain state (e.g., (Rosenthal et al., 2021; Song et al., 2018)). More recent work has analyzed the spatiotemporal features of the evoked response in relation to sensory experience (Bermudez-Contreras et al., 2023). The cortical sensory response is composed of an early transient response that reflects stimulus features and a later and slower activation (i.e., the late response) that has been causally linked to stimulus perception (Sachidhanandam et al., 2013). Bermudez-Contreras and colleagues found that repeated sensory stimulation selectively alters the spatiotemporal features of the late evoked response (Bermudez-Contreras et al., 2023). Following this study, Montagni and colleagues addressed the question of whether the secondary response to peripheral stimulation was still observable under anesthesia, whether it depended on the depth of anesthesia, and which spatiotemporal features characterized its dynamics. To this end, they explored the cortical responsiveness to external stimuli in different brain states, by recording neuronal population activity in Thy1-GCaMP6f mice (Figure 6, panels A–I; (Montagni et al., 2024, 2021). The authors compared the cortical activity dynamics following whisker stimulation at different levels of isoflurane anesthesia (Figure 6A). The neuronal response originated from the contralateral barrel cortex and rapidly spread to retrosplenial and motor cortices. Confirming previous results (Mohajerani et al., 2010; Rosenthal et al., 2021; Song et al., 2018), the activation was bilateral involving homotopic regions of both hemispheres. Unilateral multi-whisker stimulation caused a brief, local response in awake mice, which became widespread and sustained under low-dose isoflurane. Deeper anesthesia led to shorter, more localized activation, mainly in medial associative areas (Figure 6B). Analysis of the spatiotemporal features of the stimulus-evoked activity highlighted the stability of the average response in the primary sensory area, suggesting that the cortical representation of stimulus features is stable regardless of the brain state. In contrast, the integration of sensory input across the entire dorsal cortex was strongly influenced by the brain state being finely tuned by the anesthesia level. The single-trial response was composed of two sequential activation peaks (see an example in Figure 6C)-characterized by different amplitude and time to peak depending on the level of anesthesia. As anesthesia lightened, responses became more complex (Figure 6D), and a late component emerged in the sensory-evoked activity. Importantly, the probability of the late response was significantly higher with lower levels of anesthesia (Figure 6E). Consequently, minor adjustments to anesthesia levels may result in substantial variations in the distributed cortical processing of the stimulus. As the core message of the study, the authors propose that the mechanisms underlying stimulus perception may be activated even under medium levels of anesthesia. The activity at the level of both hemispheres can be simulated by The Virtual Brain (TVB) simulations of the mouse brain, developed within the HBP (Goldman et al., 2023, 2020; Montagni et al., 2024; Tort-Colet et al., 2021). Here, a model of the calcium signal was integrated in TVB to simulate the wide-field calcium signal activity recorded in the mouse (Figure 6F). The model could simulate either asynchronous activity or slow-wave oscillations for different depths of anesthesia (Figure 6G). The oscillation was slower for deeper anesthetized states, as in the experimental data (Figure 6H). To simulate the responsiveness in these brain states, the PCI was computed. The PCI remained high for the awake-like asynchronous state, but dropped to lower values for the slow oscillations of the simulated anesthetized states (Figure 6I). To uncover the underlying mechanisms and predict propagation patterns under different brain states, Capone and collaborators recently developed a novel modeling approach (Capone et al., 2023). This tool automatically generates a high-resolution mean-field model capable of simulating activity propagation across the entire cortical hemisphere. It reliably reproduces the statistical and dynamical properties of waves observed in whole-hemisphere experimental data. Also, the model allows one to infer the connectivity between brain regions from data. To this end, in the paper, the authors used calcium imaging data recorded from the dorsal cortex of a mouse cortical hemisphere (Resta et al., 2020); available in EBRAINS KG). Through an interface in the Jupyter Lab environment, available in the EBRAINS KG (Figure 6J), it is possible to modulate the brain state by interactively changing the neuromodulation and adaptation parameters and observe in real time the emergence of different dynamical regimes in both spontaneous and perturbed conditions (Capone et al., 2021). The simulation shows a rich dynamic repertoire of spatiotemporal propagation patterns with a strong dependence on the brain state. The patterns of spontaneous activity range from spirals to classical postero-anterior and rostro-caudal waves under deep anesthesia, to the dissolution of the slow-wave patterns into an asynchronous regime into the transition to wakefulness. In response to focal stimulation, the evoked wave does not propagate (top panel) when the excitability is low, while when the excitability is high a global wave is generated (bottom panel). Interestingly an intermediate level of excitability allows for the propagation of a nontrivial wave pattern (middle panel) (Capone et al.,

2023).

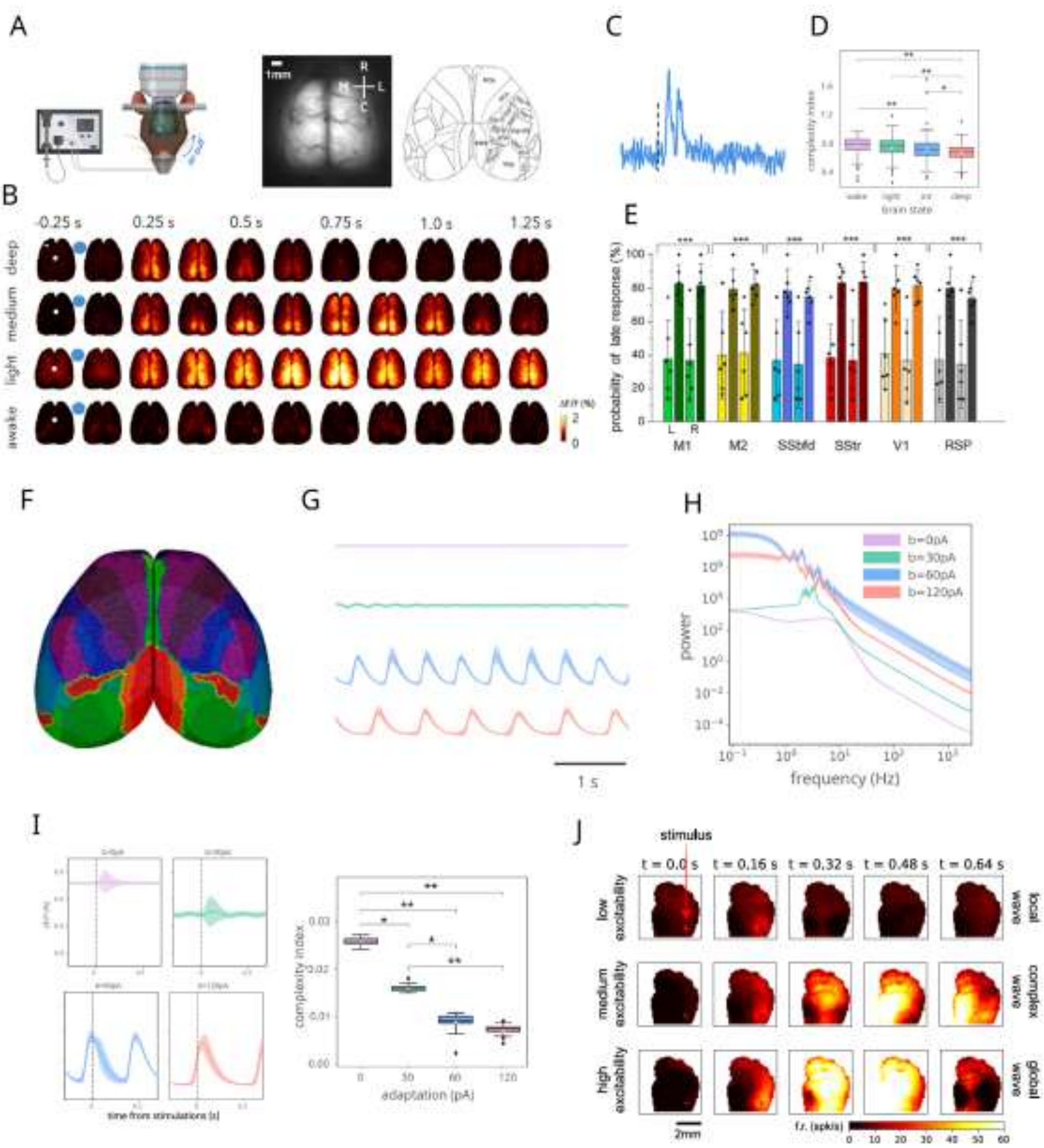

***Figure 6. Macroscale responsiveness at different brain state levels in experiments and models.*** *(A–E) Brain-state dependent probability of late stimulation response. (A) Stimulation (60 ms air puff, left); recording field of view (2 mm scale bar, middle); Allen Mouse Brain Atlas applied to parcellate among cortical areas (right). (B) Group-averaged response to sensory stimulation under deep (upper) and medium (lower) isoflurane anesthesia. Under medium anesthesia, a stronger secondary response is typically observed. (C) Exemplary sensory-evoked calcium activity under medium anesthesia showing an early and a late response. (D) PCI of sensory-evoked response, showing a decrease in complexity with increasing anesthesia levels. E. Late response probability in deep (lighter colors) and medium (darker colors) anesthesia (12 cortical regions on both hemispheres. F–I. Whole-brain model of mouse in TVB, using AdEx mean-field models. Slow oscillations induced by increasing the Spike Frequency Adaptation value (parameter b); at low SFA the network expresses an asynchronous irregular regime. (F) Calcium signals in the mouse TVB model. (G) Calcium signal simulation for different anesthesia depths (top graph: asynchronous wake-like activity). (H) Power spectra of TVB signals. (I) response to stimulation (left) and PCI Estimation with focal stimulation (right). J. Interactive simulation of state-dependent spontaneous and evoked*

*waves. (J) Propagation patterns evoked by stimulation in a mean-field mouse model inferred from calcium signals. The same network generates rich, state-dependent repertoires of spontaneous and evoked wave propagation patterns. Dorsal view of the cortical hemisphere model (pixel size 100-µm, 25 mm² field of view). EBRAINS LINK: https://wiki.ebrains.eu/bin/view/Collabs/interactive-exploration-of-brain-states.*

At the whole-brain level, a defining feature of spontaneous activity during deep NREM sleep and anesthesia is the appearance of EEG slow waves associated with neuronal OFF-periods or Down states (Steriade et al., 1993). It has been proposed that the occurrence of local OFF-periods can disrupt long-range information transmission, impairing global brain function (Lewis et al., 2014). However, intracellular recordings in cats revealed that even in deep NREM sleep, cortical neurons often remain in awake-like Up states, with OFF-periods occurring only intermittently (Chauvette et al., 2010). In humans, LFP and MUA recordings showed that slow waves and OFF-periods are predominantly local, and some brain regions may remain functionally “awake” during sleep (Hangya et al., 2011; Nir et al., 2011; Nobili et al., 2011). Moreover, depolarized phases of the slow oscillation in NREM sleep resemble fragments of wakefulness (Compte et al., 2008; Destexhe et al., 1999, 2007), suggesting that local silent periods alone may not be sufficient to impair internal communication. To address this, perturbational approaches such as TMS-EEG have been used (Ferrarelli et al., 2010; Massimini et al., 2009, 2005; Sarasso et al., 2015).

During wakefulness, TMS triggers complex, recurrent cortical responses and large-scale interactions, as measured by EEG and metrics like PCI (Casali et al., 2013; Comolatti et al., 2019). In contrast, during NREM sleep and anesthesia, the response becomes stereotypical, spatially restricted, and similar to spontaneous slow waves (Ferrarelli et al., 2010; Massimini et al., 2005; Rosanova et al., 2018; Sarasso et al., 2017). This shift is attributed to cortical bistability—i.e., the tendency of cortical circuits to enter a silent OFF-period after activation (Massimini et al., 2005). Evidence shows that stimulation during NREM induces an OFF-period—marked by a slow negative wave and high-frequency suppression—after which activity resumes in a non-causal, stochastic manner (Pigorini et al., 2015). These findings were replicated using hd-EEG and SPES in humans (Comolatti et al., 2025) and ECoG in rodents, enabling direct cross-species comparisons (Arena et al., 2021, Figure 7A). In a similar framework, Dasilva et al. (2021) showed that PCI-based cortical complexity decreases with increasing anesthetic depth, using multielectrode recordings in mice (Figure 7B).

Overall, this body of evidence suggests that cortical bistability—marked by transitions into OFF-periods—may underlie the breakdown of complex network activity during sleep, anesthesia, and potentially in pathological conditions, without requiring structural damage. In this context, Rosanova et al. (Rosanova et al., 2018) used TMS-EEG in both healthy subjects during NREM sleep and in Unresponsive Wakefulness Syndrome (UWS) patients. They found that UWS patients, despite having preserved cortical anatomy, showed sleep-like bistable EEG responses to TMS—simple waveforms followed by Down states and a loss of causal interactions and complexity (as assessed by phase-locking and PCI), similar to patterns seen during NREM. These pathological bistable dynamics were also shown to impair brain function after stroke in further TMS-EEG studies (Sarasso et al., 2021; Tscherpel et al., 2020).

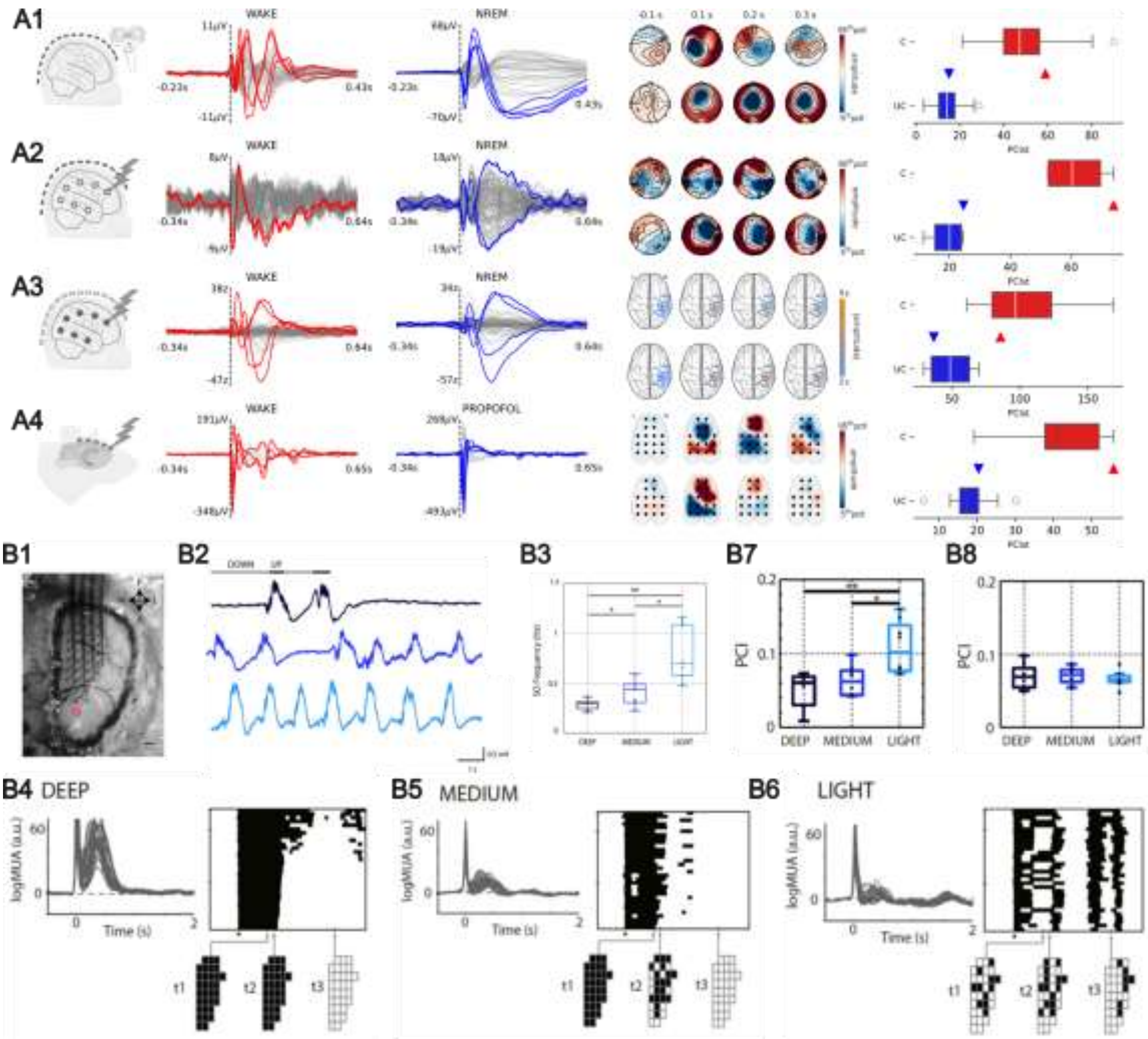

***Figure 7: Brain responsiveness across scales and species***. *(**A1-A4**) Live, data from (Arena et al., 2021; Casarotto et al., 2016); Comolatti et al., 2025). (A1) from left to right, the scalp-EEG response to TMS stimulation in human subjects in two conditions (e.g. wake vs. sleep or healthy vs. UWS), their topographical representation and the associated PCI values (triangles) with respect to a distribution obtained from a benchmark population (adapted from (Casarotto et al., 2016)). (A2,3) the scalp-EEG and the intracranial EEG (respectively) response to SPES in wakefulness and NREM sleep from the same human subject, their topographical representation and the associated PCI values (triangles) with respect to a distribution obtained from a benchmark population (adapted from Pigorini et al. in prep). (**A4**) PCI measurements in rodents. Intracranial EEG responses to perturbations by brief electrical stimulation from an intracortical electrode in area M2, during wakefulness and general anesthesia (propofol) in the same rat, their topographical representation, and the associated PCI-ST values (triangles) (adapted from Arena et al., 2021). (**B1**) Spontaneous local field potential activity from mice was recorded during SO with a superficial 32-channel multielectrode array (MEA) placed on the cortical surface (scale 500 µm). The pink circle indicates the location of the stimulation electrode. (**B2**) shows three representative recordings carried out at three different levels of anesthesia (deep, in dark blue, medium in blue, cyan in light blue respectively, color coding consistent in this panel). (**B3**) shows the average frequency of the SO. (**B4-6**) show the spatiotemporal MUA responses of all 32 channels (left) and the binary matrix (right) during the first 2s after stimulus onset during three different anesthesia levels (respectively, deep, medium and light). The spatial profile is shown at the bottom of each matrix on a visual representation of the recording MEA at three different time points (t1 to t3). Overall magnitude of perturbational complexity values under evoked (**B7**); Friedman p=0.0046; Wilcoxon Deep-Mid p=0.19, Deep-Light p=0.0078, Mid-Deep p=0.15 and spontaneous (**B8**); Friedman p=0.19 conditions in our population of mice. **B1-B8** adapted from (Dasilva et al., 2021).*

Still at the whole-brain level, computational models can capture various aspects of brain responsiveness, by providing a mathematical lens to translate anatomical structures into global dynamical states. These models share a common baseline architecture: a structural backbone, the connectome, derived from human diffusion-weighted imaging (DWI) defines the edges, while individual nodes represent parcelated brain regions (Figure 8A, left). The nodes are then equipped with a neuronal mass model representing the dynamics of a brain region of interest (Figure 8A, right). The choice of the neural mass model equipped at each node dictates what types of mechanistic questions can be addressed, trade-offs in biological realism, and how the results must be interpreted (for a recent review on virtual brains see Hashemi et al. 2025). We contrast three prominent models, ordered from purely phenomenological to highly biophysically informed: the Hopf model (Ipiña et al., 2020; Deco et al. 2017), the Montbrio-Pazo-Roxin (MPR) model (Montbrió et al., 2015) and the AdEx mean-field model (Zerlaut et al., 2018). The Hopf model is a phenomenological Stuart-Landau non-linear oscillator which is the simplest system capable of reproducing the fully synchronous and stable asynchronous states. It is an abstraction utilized to study general network properties from a dynamical viewpoint, such as global network stability and critical transitions. It tells us if a network is operating near a critical boundary, but cannot provide cellular or synaptic explanations for why it is doing so. The AdEx model is a mean-field model derived from two-populations: regular-spiking neurons with spike-frequency adaptation (excitatory) and fast-spiking neurons without adaptation (inhibitory). By keeping the link to the microscopic aspects (such as synaptic receptors), this model is built for deep mechanistic decoding. It enables direct interpretation of how molecular and chemical changes—such as anesthetics acting on synaptic receptors or neuromodulatory shifts in calcium signaling—lead to the emergence of changes at the level of brain activity (Sacha et al., 2025).  This type of model can thus relate the changes at the level of synaptic receptors, such as drugs, to effects at the level of the whole brain, such as changes in brain state induced by drugs and their associated responsiveness (PCI).  The MPR model sacrifices some of the biophysical mechanisms on the neuronal level to offer an exact macroscopic collective behaviour of a single population of quadratic integrate-and-fire neurons. It was parameterized such that each node is in a bistable regime in the absence of external input (Montbrió et al., 2015; Rabuffo et al., 2021), and as a such it is optimized for tracking collective spontaneous transitions (e.g., avalanches, bursting capacity) and relating fluid resting-state behavior directly to perturbational responsiveness (PCI). It allows researchers to interpret how baseline network "fluidity" shifts the impact of an incoming stimulus.

The Hopf model was shown to reproduce functional connectivity of resting state neuroimaging data in three distinct dynamical regimes: noise (Ghosh et al., 2008; Messé et al., 2014), fluctuating (Deco, Kringelbach, et al., 2017; Deco & Jirsa, 2012; Ghosh et al., 2008; Hansen et al., 2015) and oscillatory (Cabral et al., 2014). To further characterize the underlying dynamics, Sanz Perl et al. (2022) adopted a perturbational approach. The Hopf whole-brain model was fitted to the empirical resting state fMRI data both in the fluctuating and oscillating regimes (Figure 8B,C) and a periodic perturbation was applied to pairs of homotopic nodes (Figure 8B,C; Sanz Perl et al., 2022). The authors then computed the complexity elicited after the perturbation and also the complexity before the perturbation as a complexity level of the background signal and defined relative complexity as differences between the elicited and background complexity. Varying stimulus strength and target nodes in the subcritical regime elicited different responses in the fluctuating regime, but almost no response in the oscillatory regime. This is in line with previous results positioning the healthy brain dynamics near critical phase transition (Haimovici et al., 2013), with a departure from critical dynamics during unconsciousness (Tagliazucchi et al., 2015).

Employing a more biologically informed neural mass model, the experiments on PCI during awake vs. sleep or anesthesia could be replicated for the human brain (Goldman et al., 2023, 2020;

Sacha et al., 2025) or mice brain (Montagni et al., 2024). The mechanism to model brain state changes was based on mimicking the action of neuromodulators (McCormick, 1992) on spike-frequency adaptation, or by modeling the action of anesthetics on synaptic receptors (Sacha et al., 2025). In all cases, the switch of the brain state to slow-wave activity was accompanied by a decrease in PCI, as found in the experiments. Figure 8F illustrates that behavior in a whole-brain model using AdEx mean-fields. For low levels of adaptation, the system is in an asynchronous (fluctuating), wake-like regime, whereas for high adaptation the system exhibits synchronous sleep-like dynamics marked by slow oscillations. The response to the stimulus in the wake and sleep-like regimes differs both in complexity and variance of the temporal profile (Figure 8E; black line marks the mean over 40 realizations) and the spatiotemporal pattern (Figure 8D; color codes for time). This is consistent with the results from the simpler Hopf model, which operates on a slow timescale only, while adding the fast timescale response to the single-pulse stimuli. In the AdEx model, the complexity of the response to stimulus is highest in a fluctuating (asynchronous, awake-like) state, and lower when the model switches to slow-wave regimes with higher adaptation or modeling anesthesia. This allowed them to make a direct link to established results on decreased neuronal adaptation in the awake state, as a result of neuromodulation (Jones, 2003) and disrupted calcium signaling (Bojarskaite et al., 2020).

Another angle motivated by modeling results is how the responsiveness relates to the baseline spontaneous activity. The third model explores the relationship of the complexity of the stimulus-response to complexity-based measures of spontaneous activity (fluidity, bursting capacity, repertoire). Breyton et al. demonstrated that the working point of the whole-brain model with the highest PCI corresponds to the working point of highest spontaneous fluidity (Breyton et al., 2024). Based on this observation, they propose fluidity measures applicable to resting state EEG with comparable predictive power to PCI. The MPR model was systematically explored in the parameter space spanned by global coupling scaling (G) and noise variance (σ) both in the resting state and under stimulation. In the resting state the system exhibits maximal fluidity for intermediate values of the G and σ (Figure 8G left; (Fousek et al., 2024), where the spontaneous fluctuations give rise to rich and recurrent dynamics as captured by the dynamical functional connectivity. The working point for maximal responsiveness is close, having a slightly decreased level of intrinsic noise (Figure 8G right; (Breyton et al., 2024)) that balances out the additional external input of the stimulus. Both the slow spontaneous fluctuations and the fast stimulus-response are in this model supported by cascades on the fast time-scale—clusters of avalanches of activity (Rabuffo et al., 2021). Following these observations in the model, the authors then show on an EEG dataset (Sarasso et al., 2015) Colombo et al., 2019) that the measures of spontaneous activity capturing the fluidity in the delta band computed as circular correlation in a sliding 3 s window, the complexity (Lempel-Ziv), the bursting capacity (measure related to the avalanches), and the size of the dynamical repertoire, differentiate clearly within subjects the different states of consciousness (wake-rest/anesthesia) (Figure 8H), and at the group level perform comparably to PCI.

In summary, when addressing the whole-brain responsiveness, the brain network modeling provides a way to investigate the role of structural connectivity (Fousek et al., 2024), spatial gradients of brain organization, or the brain dynamical states (or regimes) in the shaping of the response to a stimulus. In this context, rethinking the spatiotemporal complexity metric to explicitly account for the space, (Iraji et al., 2020) will be important for future studies of complexity and responsiveness at the whole-brain scale. These could reveal differences in complexity metrics over spatial hierarchies (Wang 2020). Here the gradient of synaptic excitation can give rise to multiple temporal hierarchies (Chaudhuri et al., 2015), and the spatiotemporal structure itself can be responsible for different resonances (Petkoski & Jirsa, 2022) and stimulus propagation patterns (Spiegler et al., 2016). On the other hand, the spontaneous complex choreography of functional hierarchical organization at the whole-brain level can distinguish between different tasks and rest (Deco et al., 2021). Finally, the simulation of the effect of anesthetics on synaptic receptors showed that the whole-brain activity can switch to a less responsive state with slow waves (Sacha et al., 2025) confirming that mean-field approaches are promising for the evaluation of the brain-scale emerging activity due to drugs or receptor dysfunction at the microscopic level.

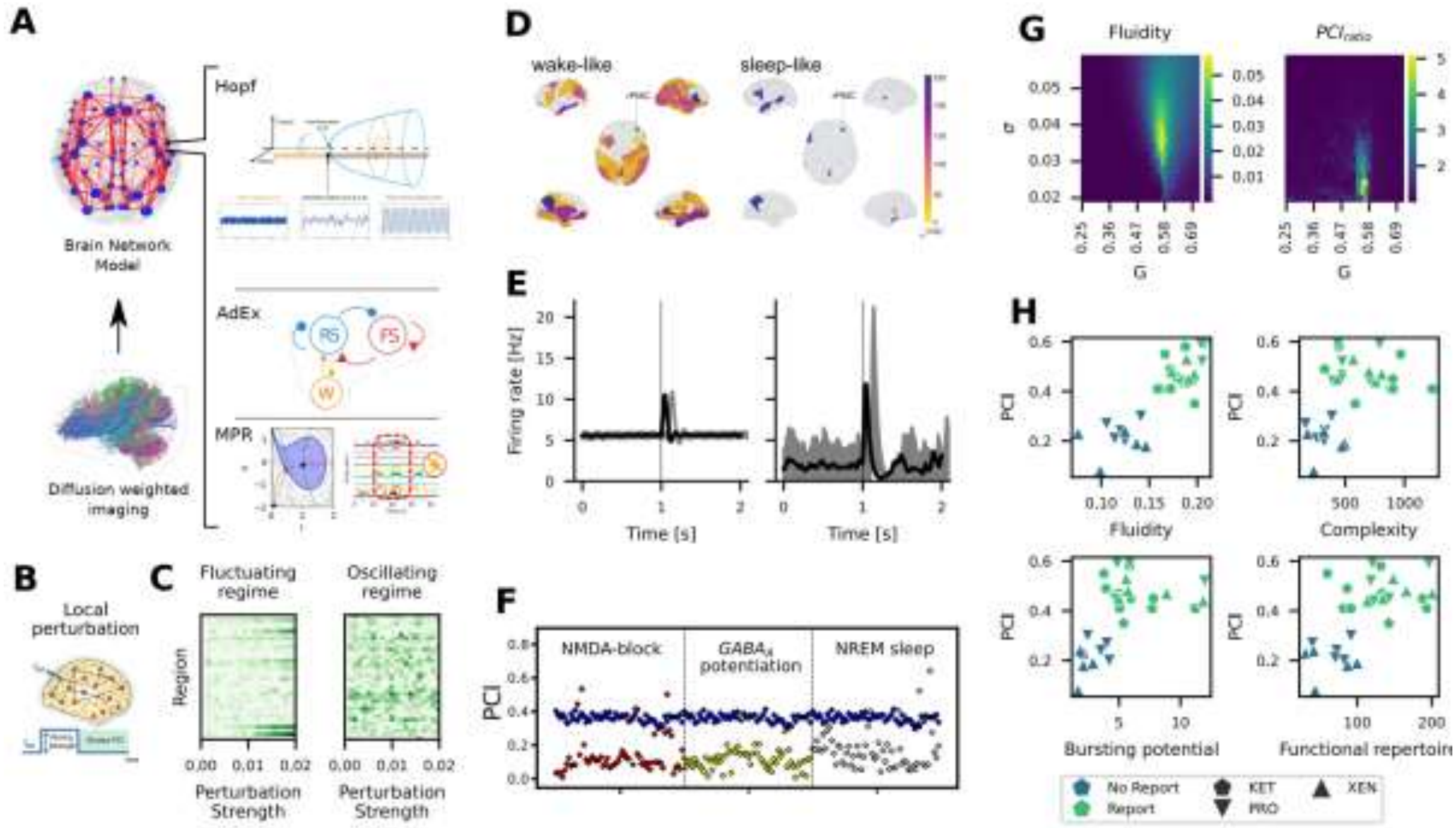

***Figure 8. Modeling responsiveness at the whole-brain level.*** *(A) The brain network model is constructed using a connectome derived from empirical DTI data, equipping the nodes with three different neural mass models to ask mechanistic questions: the phenomenological Hopf model (B,C) using response to stimulation to find support for either fluctuating (subcritical) or oscillatory (supercritical) regimes, the Montbrio Pazo Roxin Model (G,H,I) relating the dynamical features of spontaneous activity to the PCI, and the biologically most interpretable AdEx mean field (D,E,F) model investigating the role of neuronal adaptation in the complexity of the whole-brain response to stimulus. (B) The Hopf model was tuned to the sub- and super-critical regime and systematically perturbed with continuous stimulus applied with varying strength to the homotopic nodes. (C) The relative PCI after the stimulus varied across the nodes and stimulus strength across the nodes in the sub-critical regime, but there was almost no structured response in the super-critical regime. (D) Spatio-temporal propagation of the stimulus in the wake-like (low adaptation) and sleep-like (high adaptation) states, color codes time to significant deviation from baseline after the stimulus. (E) Excitatory firing rate of the stimulated brain region before and after the stimulus in the wake- and sleep-like dynamics. (F) Perturbation complexity index (PCI) of the whole-brain AdEx model in simulated anesthesia, either by NMDA block or by GABA-A potentiation, as well as in NREM sleep (high adaptation) condition. (G) For the MPR model, the working points with respect to the fluidity of spontaneous activity and Lempel-Ziv complexity of the stimulus response overlap for the parameter G that scales the network coupling, and are shifted for the noise σ. (H) Four measures of the spontaneous EEG track the PCI in the preliminary dataset of 18 subjects (left to right, top to down): fluidity computed as circular correlation in sliding 3 s window, Lempel-Ziv complexity of the z-scored and binarized signal, bursting potential, and number of unique activation patterns. C adapted from (Sanz Perl et al., 2021), F adapted from (Sacha et al., 2025); G-H adapted from (Breyton et al., 2024).*

**Box 1. Does responsiveness vary within the same brain state?**

Not only does responsiveness vary across brain states, but also within the same state. A clear example is slow oscillations.

**Up and Down States: Two States in One.**

Slow oscillations are rhythmic pattners of brain activity that occur during slow-wave sleep, deep anesthesia, and some cortical lesions (Massimini et al., 2024). These oscillations consist of two distinct phases: Up states and Down states. During Up states, brain activity resembles wakefulness, with neurons depolarizing and firing in a desynchronized, high-frequency pattern (Compte et al., 2008; Destexhe et al., 1999; 2007). Conversely, Down states are characterized by a synchronized period of neuronal silence, where membrane potentials are hyperpolarized (Steriade et al., 1993; Volgushev et al., 2006). This silent phase is key as it creates a refractory period between Up states helping to reset the network for the next Up state (Camassa et al., 2022). Down states are able to break causal interactions across cortical areas, preventing information propagation (Rosanova et al., 2018).

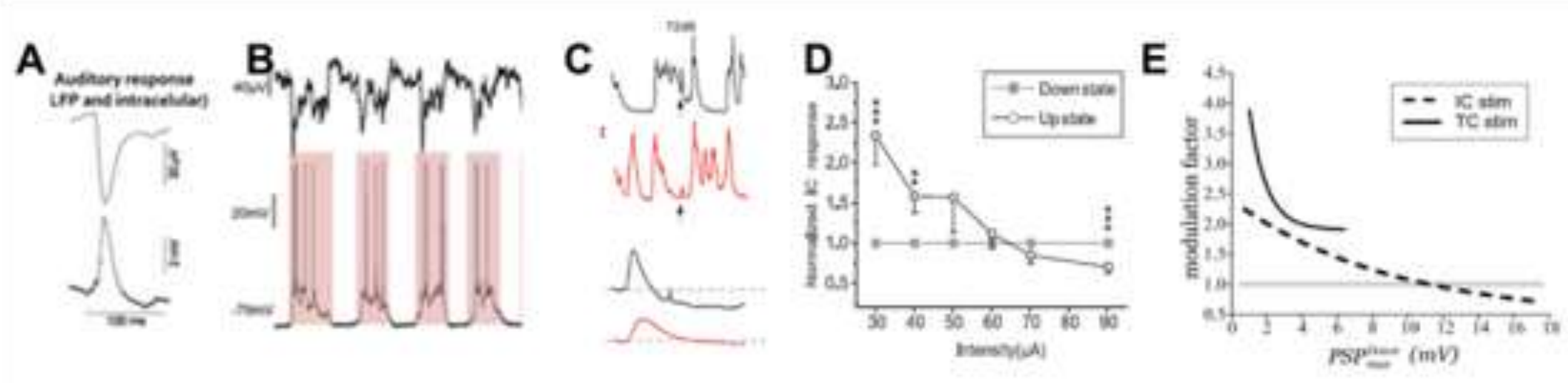

***Figure B1. State-dependent modulation of synaptic responses and model simulation.*** *(A) Auditory-evoked local field potential (LFP, top) and intracellular postsynaptic potential (PSP, bottom) recorded in rat primary auditory cortex. (B) LFP (top) and intracellular recording (bottom from auditory cortex during slow oscillations. (C) PSPs evoked by a 72 dB auditory stimulus during Down (red) and Up (black) states, illustrating the amplification of weaker inputs in the active network state. Averaged traces below. (D) Normalized PSP amplitude (relative to the Down-state response) plotted as a function of stimulus intensity for intracortical (IC) electrical stimulation. Circles, Up state; squares, Down state. Note potentiation at low intensities and attenuation at high intensities during Up states. (E) In a thalamocortical model, modulation factor (ratio of Up- to Down-state PSP amplitude) plotted against the Down-state PSP amplitude for IC (intracortical, dashed line) and thalamocortical (TC; solid line) stimulation. The TC pathway shows a larger overall gain modulation, reflecting the combined effects of thalamic and cortical network excitability. Adapted from (Reig et al., 2015).*

**Are Evoked Responses Larger in Up or Down States?**

Studies comparing responses to sensory and electrical stimuli in Up versus Down states had shown conflicting evidence: some found larger responsiveness in Up states while others in Down states. Plausible explanations exist for both extremes. In the case of Up states, these include higher excitability of the thalamocortical system -for increased responsiveness-, versus increased synaptic bombardment and thus membrane conductance (shunt), synaptic depression, or a smaller driving force for glutamatergic transmission -for decreased responsiveness-. On the other hand, Down states display lower membrane conductance, low synaptic depression -for increased responsiveness- but a larger distance to firing threshold. How can these observations be reconciled?

**Gain Modulation of Inputs in Up versus Down States**

Responsiveness depends on the intensity of stimulation, with larger stimuli evoking larger responses. However, the occurrence of Up and Down states impose a state-dependent, non-linear gain modulation depending on the intensity: low-intensity inputs are potentiated and larger inputs attenuated, resulting in a scaling or gain modulation of responses, preserving the intensity-response relationship. This effect was present for sensory (auditory) and electrical (intracortical and thalamocortical) stimulation, with intracortical activations showing the strongest evidence (Reig et al., 2015).

**Computational Models of Gain Modulation in Up vs. Down States**

A thalamocortical model of leaky integrate-and-fire neurons provided mechanisms that support the change in responsiveness in Up versus Down states. Up states amplify the synaptic input using depolarization and current fluctuations, while Down states use low conductance to generate strong postsynaptic responses. The combined effect of synaptic noise and feedforward arrangement of the network increases the detectability of small inputs and dampens large ones (see (Destexhe and Contreras, 2006; Reig et al., 2015).

**Box 2. The Perturbational Complexity Index (PCI)**

The Perturbational Complexity Index (PCI), introduced by Casali et al. (2013), is an empirical measure of brain complexity that quantifies the information content in the brain's deterministic response to a direct cortical perturbation. Originally, PCI is computed as follows (see Figure B2):

i. High-density EEG (hd-EEG) records the brain's average response (0–300 ms) to >150 Transcranial Magnetic Stimulation (TMS) pulses;
ii. Source modeling and nonparametric statistics generate a binary matrix of significant sources [SS(x,t)], reflecting the spatiotemporal pattern evoked by TMS;
iii. Lempel-Ziv complexity compresses this matrix to quantify its information content;
iv. This complexity is normalized by the source entropy of SS(x,t).

Thus, PCI is the normalized Lempel-Ziv complexity of the brain's deterministic activation pattern in response to direct cortical stimulation. This version is now referred to as PCIlz. It has been validated in 150 healthy and communicative brain-injured individuals across states like wakefulness, deep sleep, dreaming (Massimini et al., 2005), and anesthesia (Sarasso et al., 2015). An optimal cutoff (PCIlz = 0.31) was identified using ROC analysis, achieving ~100% accuracy in distinguishing conscious from unconscious states.

PCI has also been used in noncommunicative patients with disorders of consciousness, providing a reliable stratification independent of behavior (Casarotto et al., 2016; Sinitsyn et al., 2020). Crucially, PCIlz accuracy depends on TMS-EEG signal quality, and standards to enhance TMS-evoked potentials have been published (Casarotto et al., 2022; Russo et al., 2022).

To explore neural mechanisms of PCI across systems, PCIlz has been adapted to other recording modalities. While these versions cannot infer consciousness directly, they help examine underlying network dynamics. (Comolatti et al., 2019) introduced PCI state transitions (PCIst), estimating complexity from sparse intracerebral local field potentials without requiring source modeling. PCIst uses Principal Component Analysis and quantifies state transitions, and has been applied in humans (Zelmann et al. 2023) and rodents (Arena et al., 2021; Cavelli et al., 2023; Claar et al., 2022; Hönigsperger et al., 2024; Nilsen et al., 2024) distinguishing wakefulness from sleep/anesthesia.

At finer scales, PCIlz variants have assessed the complexity of Multi Unit Activity (MUA) responses to electrical stimulation in cell cultures (Colombi et al., 2021), cortical slices (D'Andola et al. 2018), and mice (Dasilva et al., 2021). These adaptations mainly differ in preprocessing and normalization. Similar approaches were used in large-scale in silico models in The Virtual Brain, where the main difference was the type of perturbation: periodic forces (Perl et al., 2021) or 50 ms square waves (Goldman et al., 2023).

Despite methodological differences, all approaches aim to quantify the spatiotemporal complexity of neural responses to perturbation. Thus, we refer to them collectively as PCI, regardless of the specific algorithm.

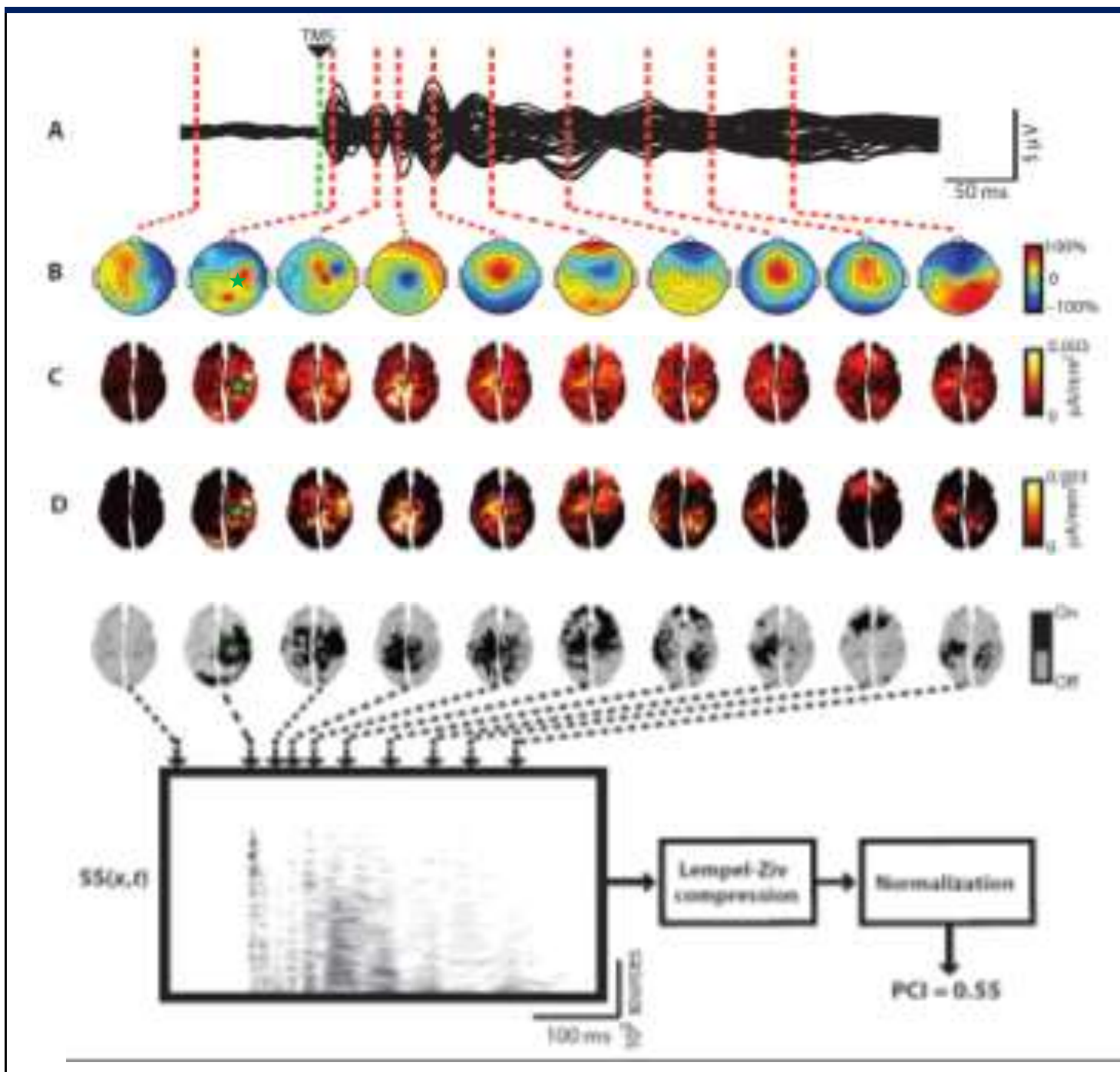

***Figure B2. (****A) Butterfly plot of 60 channels (black traces) showing averaged TMS-evoked potentials from 150 trials during wakefulness. (B) Voltage maps at selected latencies, ranging from maximum (+100%) to minimum (−100%) values. (C) A weighted minimum norm inverse solution using a three-sphere BEM model estimates cortical currents. (D) Nonparametric bootstrap statistics identify significant TMS-evoked sources. (E) A binary spatiotemporal map [SS(x,t)] is constructed, with 1 indicating significance. Sources are sorted by total activity post-stimulus. PCI is computed as the Lempel-Ziv complexity of SS, normalized by its source entropy. The TMS stimulation site is marked with a green star. Modified from (Casali et al., 2013).*

**Box 3. Microstructure of brain responses**

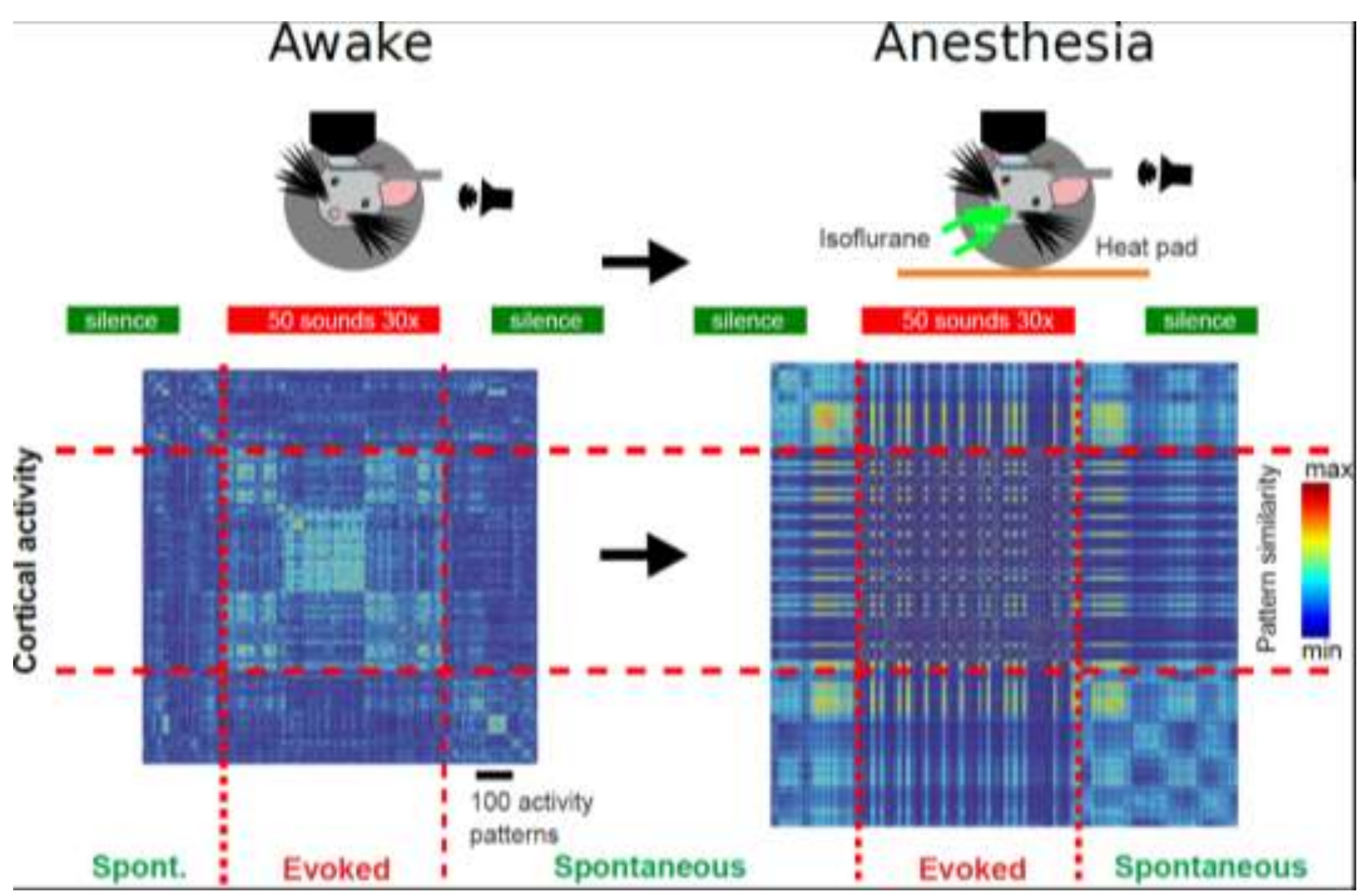

***Figure C1. Microstructure of brain responses.*** *Fine structure of spontaneous and evoked patterns of neural firing, in awake and anesthetized mice. Top: scheme of two-photon imaging of mice together with sound stimuli (50 different sounds were presented). The same population of neurons was imaged in awake and anesthetized mice. Left panels: patterns of response in awake mice. The bottom graph shows a compact representation of the correlated patterns of neural firing ("neuronal assemblies") that appear spontaneously or appear in response to the 50 different sounds. The color codes for the correlation between patterns. Right panels: same representation for anesthetized mice. Adapted from (Filipchuk et al., 2022).*

The patterns of neuronal activity can be very similar to the patterns evoked by sensory inputs. This was observed for example using VSD imaging in anesthetized cats, where it was found that spontaneous activity replays patterns of responses to sensory inputs (Kenet et al., 2003). This observation was confirmed in visual cortex of different mammals (Miller et al., 2014; Xu et al., 2012), as well as rat (Luczak et al., 2009; Sakata and Harris, 2009) and guinea-pig (Farley and Noreña, 2013) auditory cortex.

However, these studies were all done in anesthetized animals, and more recent evidence suggest that spontaneous and evoked responses are not similar in the awake brain (Filipchuk et al., 2022; Rumyantsev et al., 2020; Stringer et al., 2019). The figure illustrates that dissimilarity, in a study that was able for the first time to image the same population of neurons in the awake and anesthetized brain (here in the primary auditory cortex of mice). The correlation is high between evoked patterns and spontaneous patterns under anesthesia (right panel), for three different anesthetics, confirming previous studies in anesthetized animals. However, the corresponding correlation was almost zero in the awake animal (left panel), which shows that the evoked activity is here different than spontaneous activity. It was further shown that the evoked neuronal assemblies are specific to the different sounds, but only when the animal is awake (Filipchuk et al., 2022).

Thus, these experiments show that, although there is a response to sensory stimuli in the anesthetized animal, it is not sufficient to establish if the animal has perceived the stimulus. The data in Filipchuk et al. (Filipchuk et al., 2022) show that it is the fine structure of these responses that is important and can reveal if the sensory stimulus has been perceived by the animal.

## Conclusion

In this paper, we have reviewed aspects of neuronal responsiveness, from the microscale level of neurons and circuits, the mesoscale level of single brain areas, and the macroscale level of the whole brain. At the microscale, it is apparent that the circuit operating in an asynchronous mode displays the highest responsiveness, as seen in brain slices (D'Andola et al., 2018). The underlying mechanism is that the high levels of synaptic « noise » in asynchronous states set neurons in a high responsive mode, as seen in models of single neurons (Ho & Destexhe, 2000). This higher responsiveness is confirmed at mesoscale, and can be seen for example with Utah-array recordings comparing wake and anesthesia (Dwarakanath et al., 2025). Similarly, propagating waves occur in the asynchronous state in awake monkey (Muller et al., 2014), and sensory inputs evoke more propagating patterns (and higher PCI) in wakefulness with asynchronous states compared to slow-wave states of anesthesia in mice (Montagni et al., 2024). At the whole-brain scale, experiments also find that evoked responses are more complex and propagating compared to slow-wave states (Massimini et al., 2005), a situation which models can reproduce (Goldman et al., 2023; Sacha et al., 2025). Other measures, such as fluidity (Breyton et al., 2024) and reversibility (Camassa et al 2024) also point to the same conclusion. Collectively, these results show that asynchronous and irregular activity states set neurons in a high responsive mode, which in turn impacts network behavior and favors the propagation of activity as mesoscale propagating waves, or macroscale activity patterns that propagate across brain regions. It is therefore not surprising that the best correlate of conscious states is the asynchronous activity (Koch et al., 2016).